\documentclass[aps,prb,reprint,twocolumn,superscriptaddress,floatfix,nofootinbib]{revtex4-1}
\usepackage{amsmath,amssymb,color,comment}
\usepackage[makeroom]{cancel}
\usepackage[caption=false]{subfig}
\usepackage{mathrsfs}
\usepackage{graphicx}
\usepackage{subfig}
\usepackage[countmax]{subfloat}
\usepackage[english]{babel}
\usepackage[bookmarks=true,colorlinks,linkcolor=blue,urlcolor=blue,citecolor=blue]{hyperref}
\DeclareMathOperator{\Tr}{Tr}
\newcommand{\mbeq}{\overset{!}{=}}
\usepackage{bm}
\usepackage{braket}
\usepackage{dsfont}

\DeclareMathOperator\arctanh{arctanh}
\def\d{\mathrm d}

\definecolor{mypink1}{rgb}{0.858, 0.188, 0.478}

\begin{document}
\title{Concurrence of dynamical phase transitions at finite temperature in the fully connected transverse-field Ising model}

\author{Johannes Lang}
\affiliation{Physik Department, Technische Universit\"at M\"unchen, 85747 Garching, Germany}

\author{Bernhard Frank}
\affiliation{Physik Department, Technische Universit\"at M\"unchen, 85747 Garching, Germany}

\author{Jad C.~Halimeh}
\affiliation{Max Planck Institute for the Physics of Complex Systems, 01187 Dresden, Germany}
\affiliation{Physik Department, Technische Universit\"at M\"unchen, 85747 Garching, Germany}

\date{\today}

\begin{abstract}
We construct the finite-temperature dynamical phase diagram of the fully connected transverse-field Ising model from the vantage point of two disparate concepts of dynamical criticality. An analytical derivation of the classical dynamics and exact diagonalization simulations are used to study the dynamics after a quantum quench in the system prepared in a thermal equilibrium state. The different dynamical phases characterized by the type of non-analyticities that emerge in an appropriately defined Loschmidt-echo return rate directly correspond to the dynamical phases determined by the spontaneous breaking of $\mathbb{Z}_2$ symmetry in the long-time steady state. The dynamical phase diagram is qualitatively different depending on whether the initial thermal state is ferromagnetic or paramagnetic. Whereas the former leads to a dynamical phase diagram that can be directly related to its equilibrium counterpart, the latter gives rise to a divergent dynamical critical temperature at vanishing final transverse-field strength.
\end{abstract}
\maketitle

\section{Introduction}
Phase transitions are a textbook subject of condensed-matter and statistical physics. They are ubiquitous in daily life as well as the topic of intense investigation in experimental setups in physics laboratories around the globe. By varying a set of parameters, usually the temperature or the strength of an external field (pressure), the system settles in different equilibrium phases, which are not adiabatically connected. Equivalently, the Gibbs free energy is a non-analytic function of this set of parameters, even though the Hamiltonian describing the system is fully analytic with no singularities whatsoever. From the perspective of Landau theory, phase transitions involve the spontaneous breaking of a symmetry in the equilibrium state of the system from one phase to another distinguished by a local order parameter. \cite{Cardy,Sachdev,Ma} The framework of Wilson's renormalization group \cite{Fisher1974} utilizes scale invariance and transformations as a powerful tool to study equilibrium criticality. Equilibrium phase transitions have also been detected \cite{Greiner2002} in quantum ultracold-gas setups, and they have been studied extensively in this context both theoretically and experimentally.\cite{Bloch2008}

Furthermore, with the high degree of experimental control in ultracold-gas and condensed-matter setups,\cite{Hart2014,Suda2015,Mitrano2016} probing the out-of-equilibrium dynamics of quantum many-body systems has become a real possibility. A natural point of interest in such experiments is the concept of dynamical phase transitions (DPT) and how they relate to their equilibrium counterparts in quantum many-body systems. DPT arise in the dynamics following a quench in a certain control parameter in the Hamiltonian of the system. In principle, DPT fall into two main definitions or types:\cite{Zvyagin2017} The first, DPT-I, \cite{Moeckel2008,Eckstein2008,Eckstein2009,Moeckel2010,Sciolla2010,
Sciolla2011,Gambassi2011,Tsuji2013,Sciolla2013,Chandran2013,Maraga2015,
Smacchia2015,Langen2016,Marcuzzi2016,Zunkovic2016a,
Halimeh2016a,Halimeh2016b,Homrighausen2017,Halimeh2017a,Zunkovic2016b} relies on the system reaching a steady state, from which a local order parameter can be extracted, before the system settles into thermal equilibrium. DPT-I is a Landau-type transition that depends on the long-time average of the order parameter, whereby if this average is zero (finite), then the system is in a disordered (ordered) steady state. The final value of the quench parameter that separates between these two phases for a given initial condition is the DPT-I (dynamical) critical point. DPT-I has recently been investigated experimentally with trapped ions.\cite{Zhang2017} A theoretical review of the DPT-I can be found in Ref.~\onlinecite{Mori2017}. The second type, DPT-II, does not rely on a local order parameter or on the system settling into a steady state, but rather on non-analyticities in the form of cusps in the Loschmidt-echo return rate,\cite{Heyl2013} which is a dynamical analog of the equilibrium (boundary) free energy. The final value of the quench control parameter, for a given initial condition, would then give rise to different phases, each characterized by its own kind of cusps or lack thereof.\cite{Heyl2013,Halimeh2016b,Halimeh2017a,Homrighausen2017}
\begin{figure}[t]
\centering
\includegraphics[width=.95\columnwidth]{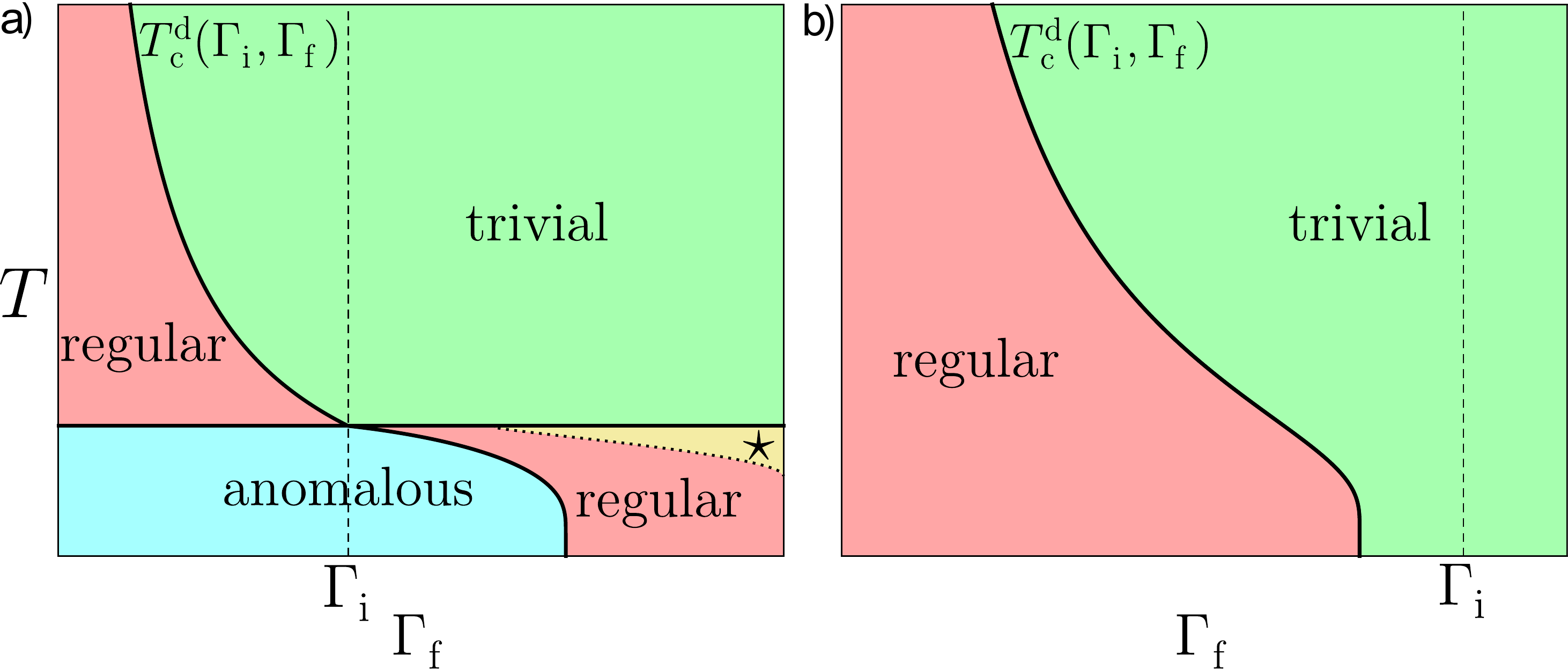}
\caption{(Color online) Finite-temperature dynamical phase diagram of the fully connected transverse-field Ising model. In (a) the system is initialized with $\Gamma_\text{i}<\Gamma_\text{c}^\text{e}(T=0)$ whereas the quench in (b) starts always in a paramagnetic state. The full lines indicating the critical lines are analytical results. The dotted line separates the area where the Loschmidt-echo return rate exhibits a thermal cutoff ($\star$) from the rest of the regular phase, but this is not a separate phase in itself and is still within the latter.}
\label{fig:phasediagram} 
\end{figure}
In general, the critical value of the quenching parameter separating the different phases of the DPT-II is not equal to the equilibrium critical point,\cite{Andraschko2014,Vajna2014} although in certain special cases, such as quenching from the fully disordered ground state of the one-dimensional transverse-field Ising model with power-law interactions,\cite{Heyl2013,Halimeh2016b} the dynamical and equilibrium critical points are the same. Since the seminal work\cite{Heyl2013} introducing it, the DPT-II has been extensively studied theoretically,\cite{Karrasch2013,Divakaran2016,Budich2016,Karrasch2017,Mera2017,Sirker2017,Zhou2017,Zunkovic2016a,Zunkovic2016b,Halimeh2016a,Homrighausen2017,Andraschko2014,Vajna2014,Heyl2014,Heyl2015,Abeling2016,Halimeh2016b,Dutta2017,Heyl2017,Halimeh2017a,Weidinger2017} and also recently experimentally observed in setups of ultracold atoms in optical lattices\cite{Flaeschner2016} and ion traps.\cite{Jurcevic2017} A review of the recent theoretical and experimental progress on DPT-II is provided in Ref.~\onlinecite{HeylReview}.

The relationship of the DPT-I and DPT-II has also been numerically studied in the nonintegrable long-range (power-law interacting $\propto1/r^\alpha$ with inter-spin distance $r$ and interaction exponent $\alpha$) transverse-field Ising chain\cite{Zunkovic2016b,Halimeh2016b,Halimeh2017a} and the fully connected transverse-field Ising model (FC-TFIM),\cite{Homrighausen2017} which is the $\alpha=0$ limit of the former. Within the numerical precision of these studies, it seems that the DPT-I and DPT-II may overlap for sufficiently long-range interactions ($\alpha\lesssim2.4$). For short-range interactions, on the other hand, the existence of a DPT-I is not confirmed,\cite{Halimeh2016a} and a comparison is, therefore, not possible.

Despite the vast amount of work on the DPT-I and DPT-II, the initial state has in most cases been the ground state of the pre-quench Hamiltonian, although finite-temperature dynamical phase transitions have recently been investigated in the case of the short-range Ising\cite{Abeling2016} and Kitaev chains,\cite{Sirker2017} where it has been shown that at any finite temperature $T>0$ the return rate is completely smooth for a quench that would lead to cusps at $T=0$. Additionally, DPT-II due to quenches involving excited pure initial states\cite{Andraschko2014,Wang2017} has also been studied. In this work, we investigate the finite-temperature behavior of both the DPT-I and DPT-II in the FC-TFIM, and construct the corresponding dynamical phase diagram Fig.~\ref{fig:phasediagram} that we show is common to both. The infinite-range interactions inherent to the FC-TFIM make it a particularly interesting platform to investigate the relation at finite temperature between these two concepts of DPT, mainly because such interactions lead to a finite-temperature equilibrium phase transition, allowing the DPT-I to exhibit rich behavior that is not present in short-range models in one dimension.\cite{Halimeh2016a} Additionally, the full-connectedness of the model allows for a mean-field treatment and also renders it integrable, which makes exact diagonalization (ED) of very large system sizes tractable. Moreover, the integrability class of the FC-TFIM is less restrictive than the quadratic fermionic Hamiltonians characteristic of the short-range models that have been the focus of previous investigations in DPT-II at finite temperature.

The rest of the paper is organized as follows: In Sec.~\ref{sec:Model} we present the FC-TFIM and the quench protocols that we adopt to construct the phase diagram for different initial conditions. In Sec.~\ref{sec:classical} we investigate the classical dynamics of our system and derive the critical line of the DPT-I for ferromagnetic and paramagnetic initial conditions. In Sec.~\ref{sec:Loschmidt} we review the interferometric generalization of the Loschmidt-echo return rate for thermal systems, and show why it is not a suitable return rate for the integrability class of the FC-TFIM. Consequently, we introduce in Sec.~\ref{sec:rFid} the more appropriate, but difficult-to-compute fidelity Loschmidt echo, and finally in Sec.~\ref{sec:rtilde} a proper and efficient form of the return rate that exclusively probes the dynamical criticality introduced by the quench. The main numerical results of our work are presented in Sec.~\ref{sec:results} where, using ED, we calculate the Loschmidt-echo return rate for various quantum quenches at various preparation temperatures to elucidate the construction of the dynamical phase diagram shown in Fig.~\ref{fig:phasediagram}. We conclude in Sec.~\ref{sec:conclusion}.

\section{Model and quench}\label{sec:Model}
In this work we probe the finite-temperature dynamical phase diagram of the FC-TFIM described by the Hamiltonian

\begin{align}\label{eq:FC-TFIM}
H(\Gamma)=-\frac{J}{2N}\sum_{i\neq j=1}^Ns^z_is^z_j-\Gamma\sum_{j=1}^Ns^x_j-\Lambda\sum_{j=1}^Ns^z_j\;,
\end{align}

\noindent with (ferromagnetic) coupling constant $J>0$ and system size $N$. The Kac normalization factor \cite{Kac1963} $1/N$ in the interaction term is to ensure energy extensivity in the thermodynamic limit. $s^{z(x)}_j$ is the projection of the spin operator of site $j$ onto the $z$ ($x$) direction. For quenches starting in a paramagnetic state the small longitudinal field proportional to $\Lambda\ll\Gamma$ is needed to allow symmetry breaking in case of an ordered long-time steady state. This procedure will be described in more detail in Sec.~\ref{sec:results}. Thus, unless otherwise specified, we set $\Lambda=0$. The result is the usual FC-TFIM, which has a finite-temperature equilibrium phase diagram\cite{Das2006,Wilms2011} with ordered and disordered phases separated by the equilibrium critical line

\begin{align}\label{eq:CriticalLine}
T_{\text{c}}^{\text{e}}(\Gamma)=\Gamma\left[\ln\left(\frac{J+2\Gamma}{J-2\Gamma}\right)\right]^{-1},
\end{align}

\noindent with zero-field thermal critical point $T_{\text{c}}^{\text{e}}|_{\Gamma\to0}=\lim_{\Gamma\to0}T_{\text{c}}^{\text{e}}(\Gamma)=J/4$, and zero-temperature quantum critical point $\Gamma^{\text{e}}_{\text{c}}(T=0)=J/2$, where $\Gamma^{\text{e}}_{\text{c}}(T)$ is the inverse of $T_{\text{c}}^{\text{e}}(\Gamma)$. Our notation shall entail referring to $\Gamma^{\text{e}}_{\text{c}}(T)$ as the quantum critical point at temperature $T$, and to $T_{\text{c}}^{\text{e}}(\Gamma)$ as the thermal critical point at transverse-field strength $\Gamma$.

We note here that the FC-TFIM is a special case of the Lipkin-Meshkov-Glick (LMG) model, which was devised by the eponymous scientists for shape phase transitions in nuclei.\cite{Lipkin1965,Meshkov1965,Glick1965} The LMG model describes the infinite-range anisotropic XY model in a transverse magnetic field along the $z$ direction. Despite its simple structure, it exhibits a finite-temperature phase transition,\cite{Botet1982,Botet1983} and can be experimentally realized in quantum-optical settings.\cite{Morrison2008,Neyenhuis2017,Zhang2017,Bernien2017,Hess2017} Since its introduction, it has been extensively studied in statistical physics, and even its spectrum in the thermodynamic limit has been analytically calculated in Ref.~\onlinecite{Ribeiro2008}.

We prepare our system at temperature $T=1/\beta$ in the thermal state

\begin{align}
\rho_{\text{i}}=\frac{\text{e}^{-\beta \mathcal{H}(\Gamma_{\text{i}})}}{\Tr \text{e}^{-\beta \mathcal{H}(\Gamma_{\text{i}})}}\;,
\end{align}

\noindent and set the Boltzmann and Planck constants as well as $J$ to unity throughout the manuscript. Here, $\mathcal{H}$ refers to the full (mean-field) Hamiltonian when $\rho_{\text{i}}$ is in the paramagnetic (ferromagnetic) equilibrium phase and we define the partition function as $Z=\Tr \exp{\left(-\beta \mathcal{H}(\Gamma_{\text{i}})\right)}$. Whereas the full Hamiltonian is given in~\eqref{eq:FC-TFIM}, the mean-field Hamiltonian is

\begin{align}
H_{\text{MF}}=\sum_{i=1}^N \left(m s^z_i-\Gamma s_i^x\right)\;,
\end{align}
with the equilibrium mean-field magnetization $m=\sum_i\langle s^z_i\rangle/N$ obtained self-consistently by solving
\begin{align}
2\sqrt{\Gamma^2+m^2}=\tanh{\left(\frac{\beta}{2}\sqrt{\Gamma^2+m^2}\right)}\;.
\end{align}
This procedure is necessary to enforce a finite magnetization in the ferromagnetic initial state.

At time $t=0$, we abruptly switch the transverse-field strength from $\Gamma_{\text{i}}$ to $\Gamma_{\text{f}}\neq\Gamma_{\text{i}}$, thereby initiating the dynamics of our system, which is always propagated by the full Hamiltonian~\eqref{eq:FC-TFIM}.

\section{Classical dynamics}\label{sec:classical}
We now derive the classical dynamics of our model. One can rewrite the Hamiltonian~\eqref{eq:FC-TFIM} in the form

\begin{align}\label{eq:classical}
\begin{split}
H=&-\frac{1}{2 N}S_zS_z -\Gamma S_x -\Lambda S_z\\
\simeq&-\frac{S^2}{2 N}\cos^2{\theta}-\Gamma S \sin{\theta}\cos{\phi}-\Lambda S \cos{\theta}\\\equiv& H(\theta,\phi)\;,
\end{split}
\end{align}

\noindent for the total spin vector $\bm{S}=(S_x,S_y,S_z)^{\intercal}=\sum_i \langle \bm{s}_i\rangle$, with conserved spin length $S^2=S_x^2+S_y^2+S_z^2$. The first equality in~\eqref{eq:classical} is, up to an irrelevant constant, an exact reformulation of \eqref{eq:FC-TFIM}, while the second uses the classical continuous representation $\bm{S}=S\left(\cos{\phi}\sin{\theta},\sin{\phi}\sin{\theta},\cos{\theta}\right)^{\intercal}$ and thus neglects intensive modifications to the Hamiltonian arising from the non-commutativity of the spin operators. The system of coupled equations of motion for $\bm{S}$ is given by
\begin{align}
\frac{\d \bm{S}}{\d t}=-\text{i}\left[\bm{S},H\right].
\end{align}
In the classical formulation, where the commutator is replaced by the Poisson bracket, these turn into
\begin{align}
\begin{split}
\frac{\d\theta}{\d t}&=\Gamma\sin{\phi},\\
\frac{\d\phi}{\d t}&=\Gamma \cos{\phi}\cot{\theta}-\frac{S}{N}\cos{\theta}-\Lambda\;,
\end{split}
\end{align}
which show no relaxation.

The (conserved) energy of the initial state after the quench is given by

\begin{widetext}
\begin{align}
E=\frac{1}{Z}\int_0^{2\pi}\d\phi\int_0^\pi \d\theta\int_0^1 \d s\; s^2 \sin{\theta}\;\text{e}^{-\beta H_\text{i}(\theta,\phi)}D(sN/2) H_{\text{f}}(\theta,\phi)\;,
\end{align}
\end{widetext}

\noindent with $s=2S/N\in[0,1]$, and where $H_{\text{i}(\text{f})}$ corresponds to~\eqref{eq:classical} with $\Gamma=\Gamma_{\text{i}(\text{f})}$. The degeneracy factor of the subspace with fixed value of $S$ is given by

\begin{align}\label{eq:degen}
D(S)&=
\begin{pmatrix}
	N \\
	\frac{N}{2}-S
\end{pmatrix}-
\begin{pmatrix}
	N \\
	\frac{N}{2}-S-1
\end{pmatrix},
\end{align} 

\noindent and

\begin{align}
Z=\int_0^{2\pi}\d\phi\int_0^\pi \d\theta\int_0^1 \d s\; s^2 \sin{\theta}\;\text{e}^{-\beta H_\text{i}(\theta,\phi)}D(sN/2)
\end{align}

\noindent is the partition function. The limit $N\to\infty$ allows for a saddle-point expansion around the maximum of the product $D(sN/2)\exp{\left(-\beta H_{\text{i}}(\theta,\phi)\right)}$, which fixes $s=\bar{s}$, $\theta=\bar{\theta}$, and for $\Gamma_{\text{i}}\neq0$ also $\phi$ to exact values. Thermal fluctuations around these values are suppressed by factors of $\exp{(-\beta N)}$ and can thus be neglected. This implies that the partition function and all thermal expectation values are simply given by the evaluation at the saddle point of $D(sN/2)\exp{\left(-\beta H_{\text{i}}(\theta,\phi)\right)}$. Consequently, one finds

\begin{align}
E=-N\left[\frac{\bar{s}^2}{8}+\Gamma_{\text{i}}\left(\Gamma_{\text{f}}-\frac{1}{2}\Gamma_{\text{i}}\right)\right].
\end{align} 

\noindent The long-time-averaged magnetization in the $z$-direction can vanish only if the classical spin vector can overcome the equator of the Bloch sphere. Thus a phase transition occurs if the initial state is prepared such that its energy after the quench is sufficiently large. This allows us to determine the critical line from the equation
\begin{align}
E(\bar{s})=-\Gamma_{\text{f}}\frac{N \bar{s}}{2}\;.
\end{align}
In the case of a $\mathbb{Z}_2$ symmetry-broken initial state, inserting the saddle-point equations
\begin{align} \label{eq:s_max}
\begin{split}
\frac{\beta \bar{s} }{4}\cos^2{\bar{\theta}}+ \frac{\beta \Gamma_\text{i}}{2}  \sin \bar{\theta}& =\arctanh{\bar{s}}, \\ \theta_{\text{i}}=\bar{\theta}&=\arcsin{\frac{2\Gamma_{\text{i}}}{\bar{s}}}\;,
\end{split}
\end{align}
 for the spin length and direction allows us to determine the analytic expression for the dynamical critical temperature valid in the thermodynamic limit $N\to\infty$, which reads
\begin{align}\label{eq:T_c_dyn}
T_{\text{c}}^{\text{d}}(\Gamma_{\text{i}},\Gamma_{\text{f}})=\frac{2\Gamma_{\text{f}}-\Gamma_{\text{i}}}{2\arctanh{(4\Gamma_{\text{f}}-2\Gamma_{\text{i}})}}\;.
\end{align}

For a $\mathbb{Z}_2$-symmetric initial state, the saddle point of the partition function is given by
\begin{align}
\bar{\theta}=\frac{\pi}{2}\;,\;\;\;\;\;\bar{\phi}=0\;,\;\;\;\;\;\frac{\beta\Gamma_{\text{i}}}{2}=\arctanh{\bar{s}}\;,
\end{align}
and the classical system shows no dynamics for $\Gamma_{\text{f}}>\Gamma_{\text{i}}$ since the system is initialized in the ground state (for fixed $s$) of both the initial and final Hamiltonians. Relaxation is only possible if the ground state of the final system at $s=\bar{s}$ is ferromagnetic. In this case $\langle S_z \rangle$ will relax to a finite value (given a sufficiently large seed). As the effective spin length $s$ is not the same as in the equilibrium phase corresponding to $\Gamma_{\text{f}}$, but rather to that of the equilibrium phase at $\Gamma_{\text{i}}$, the resulting critical final field strength is given by 

\begin{align}\label{eq:G_c_dyn}
\Gamma_{\text{c}}^{\text{d}}=\frac{\bar{s}}{2}=\frac{1}{2}\tanh\frac{\Gamma_{\text{i}}}{2T}\;,
\end{align}

\noindent for $\Gamma_{\text{i}}>\Gamma_{\text{c}}^{\text{e}}$, as immediately follows from the condition

\begin{align}
\frac{\d^2H(\theta,0)}{\d\theta^2}\bigg|_{\theta=\frac{\pi}{2}}\mbeq0\;.
\end{align}
It is important to note that the collapse of the partition function in the thermodynamic limit is also true for the quantum-mechanical treatment of the problem where it is therefore also allowed to fix $s$ according to \eqref{eq:s_max} as $N\to \infty$.

\section{Interferometric Loschmidt-echo return rate}\label{sec:Loschmidt}
There are several distinct generalizations of the ground-state Loschmidt echo for finite temperatures. One straightforward extension to thermal states is the interferometric Loschmidt amplitude, which has recently been defined as \cite{Dutta2017,Heyl2017}

\begin{align}\label{eq:LA_thermal}
G_\text{I}(t)=\Tr\big\{\rho_{\text{i}}\text{e}^{-\text{i} H(\Gamma_{\text{f}})t}\big\}\;,
\end{align}

\noindent where $H$ refers to the full Hamiltonian ~\eqref{eq:FC-TFIM}. In the limit of zero temperature this reduces to the original Loschmidt amplitude\cite{Heyl2013} $\langle\psi_{\text{i}}|\exp(-\text{i} H(\Gamma_{\text{f}})t)|\psi_{\text{i}}\rangle$, with $|\psi_{\text{i}}\rangle$ the ground state of the pre-quench Hamiltonian $H(\Gamma_{\text{i}})$. In~\eqref{eq:LA_thermal}, the evolution time $t$ takes the place of the complexified inverse temperature, making it a dynamical analog of a boundary partition function. Consequently, the corresponding dynamical analog of the thermal free energy density in equilibrium is the Loschmidt-echo return rate

\begin{align}\label{eq:LERR_thermal}
r_\text{I}(t)=-\lim_{N\to\infty}\frac{1}{N}\ln|G_\text{I}(t)|^2\;.
\end{align}

At $T=0$, the DPT-II is connected to non-analytic cusps in~\eqref{eq:LERR_thermal}.\cite{Heyl2013,Heyl2014,Heyl2015} In addition to Ref.~\onlinecite{Heyl2013} that studied the DPT-II in the integrable one-dimensional nearest-neighbor transverse-field Ising model (NN-TFIM), the DPT-II has also been investigated in the nonintegrable one-dimensional long-range \cite{Zunkovic2016b,Halimeh2016b,Halimeh2017a} and the fully connected transverse-field Ising models.\cite{Zunkovic2016a,Homrighausen2017} Unlike the DPT-I, which has only two distinct phases, the DPT-II exhibits three distinct dynamical phases.\cite{Halimeh2016b,Halimeh2017a,Homrighausen2017} Starting from an ordered ground state, quenches across a dynamical critical point give rise to regular cusps (\textit{i.e.}~cusps in every oscillation) in the Loschmidt-echo return rate. On the other hand, for quenches below this dynamical critical point, the return rate displays no cusps when the interactions are short-range, while for sufficiently long-range interactions,\cite{Halimeh2016b} a new kind of \textit{anomalous} cusps (\textit{i.e.}~cusps appearing only after a certain number of smooth oscillations) have been shown to emerge.\cite{Halimeh2016b,Homrighausen2017,Halimeh2017a} Moreover, the DPT-I and DPT-II seem to be intimately connected, at least for long-range interactions.\cite{Zunkovic2016b,Halimeh2016b,Homrighausen2017}

\begin{figure}[htp]
\centering
\includegraphics[width=.95\columnwidth]{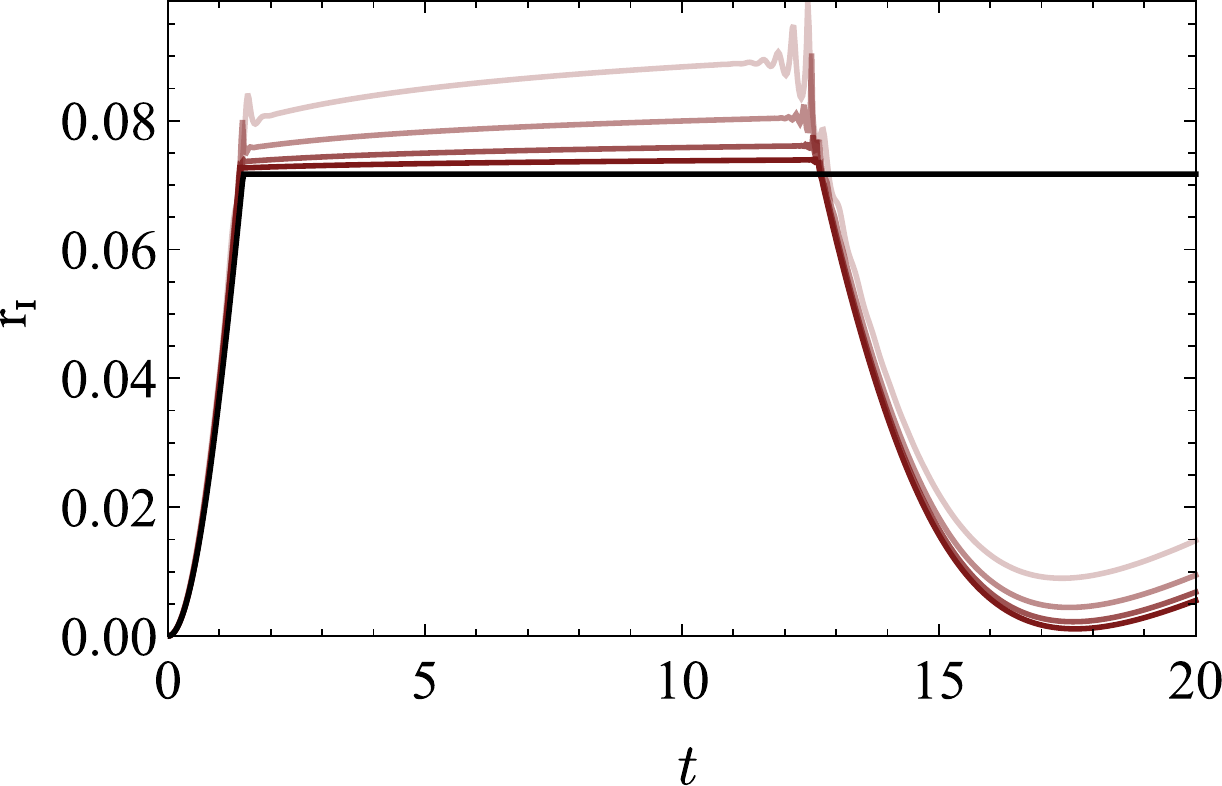}
\caption{(Color online) Comparison of exact finite-size results with the analytical expression \eqref{eq:ana}, shown in black, for the return rate~\eqref{eq:LERR_thermal} at short times with inverse temperature $\beta=5$ and a trivial quench $\Gamma_{\text{i}}=\Gamma_{\text{f}}=0$. System size from light to dark red is $N=200$, $400$, $800$, and $1600$.}
\label{fig:scaling}
\end{figure}

For the infinitely connected model, however, even trivial quenches from $\Gamma_\text{i}\to\Gamma_\text{i}$, that at $T=0$ result in $r_\text{I}(t)\equiv 0$, can show a rich non-analytic behavior of the return rate at finite temperatures. To understand this in some more detail, let us consider the easiest case without a transverse field. The associated Hamiltonian is diagonal in $S_z$ and given by
\begin{align}\label{eq:NoTF}
H=-\frac{1}{2 N}S_z^2.
\end{align}
Following the calculation outlined in Appendix~\ref{app:thermal_ana}, we obtain in the thermodynamic limit a sharp signature in the short-time return rate,
\begin{widetext}
\begin{align}\label{eq:ana}
r_\text{I}(t)=\frac{1}{4}\min{\left\{\frac{s_0\left(1-s_0\right)\left(1-2 s_0\right)^2\left[1+s_0\left(s_0-1\right)\beta\right]t^2}{\left\{1+s_0\left(s_0-1\right)\left[2\beta+s_0\left(s_0-1\right)\left(t^2+\beta^2\right)\right]\right\}},\beta-4s_0^2\beta-8\ln{\left(2-2 s_0\right)}\right\}}\;,
\end{align}
\end{widetext}
where $s_0$ solves the saddle-point condition
\begin{align}
\beta\left(\frac{1}{2}-s_0\right)=2\arctanh{\left(1-2s_0\right)}\;.
\end{align}
This result compares well with the full, numerically evaluated expression for inverse temperatures $\beta < 5.5$. Further numerical investigation shows that the sharp cutoff in the first peak survives for inverse temperatures as low as $\beta\approx5.9$, which is deep inside the ferromagnetic phase. Fig.~\ref{fig:scaling} shows a comparison of exact interferometric return rates $r_\text{I}(t)$ for finite systems with the analytical expression \eqref{eq:ana}. One clearly sees the convergence for $N\to\infty$ of the numerical data toward the analytical plateau, creating an increasingly sharp thermal cusp in the first peak in the process. In addition to the cutoff in the first peak, numerical simulations for system sizes of up to $N=2\times10^5$ show further cusps appearing at late times, reminiscent of the anomalous phase previously investigated in the FC-TFIM at $T=0$ in Refs.~\onlinecite{Halimeh2016b,Homrighausen2017}. For the trivial quench at finite temperatures, these cusps are nothing else but the Lee-Yang zeros in the complex plane of the partition function.\cite{Lee1952} We show further results for $r_\text{I}(t)$ obtained from finite quench distances at finite temperatures in Appendix~\ref{app:thermal_num}.

\section{Fidelity Loschmidt-echo return rate}\label{sec:rFid}
In a quantum quench setup one is interested in the time evolution governed by a Hamiltonian $H_{\text{f}}$, where the system is initially prepared in a thermal equilibrium state $\rho_{\text{i}}$. Under the condition that $\rho_{\text{i}}$ is not diagonal in the eigenbasis $\left\{|\Phi_{\text{f}} \rangle \right\}$ of $H_{\text{f}}$, the Loschmidt return function $r_\text{I}(t)$ can show non-analytic behavior due to the nontrivial overlap of the states making up $\rho_{\text{i}}$ with the eigenstates of $H_{\text{f}}$. However starting with a genuine density matrix and not a pure state can give rise to a type of non-smooth features in $r_\text{I}(t)$ that we are not interested in and that would already appear in a trivial quench $\Gamma_{\text{i}} \to \Gamma_{\text{i}}$, as discussed in detail in Sec.~\ref{sec:Loschmidt}.
Therefore, a particularly interesting choice of the finite-temperature Loschmidt amplitude is given by the fidelity of the initial and time-evolved density matrices\cite{Zanardi2007,Venuti2011}
\begin{align}\label{eq:LA_Fid}
G_\text{F}(t)=\Tr\sqrt{\sqrt{\rho(0)}\rho(t)\sqrt{\rho(0)}}\;.
\end{align}
As with the interferometric definition \eqref{eq:LA_thermal}, \eqref{eq:LA_Fid} also reproduces the zero-temperature Loschmidt echo.
The advantage of this definition is, however, that it forms a metric, measuring the distance over which the density matrix has evolved during the time $t$. This also implies that the corresponding fidelity return rate 
\begin{align}
r_\text{F}(t)=-\lim_{N\to\infty} \frac{2}{N}\ln{G_\text{F}(t)}
\end{align}
vanishes identically in case of any trivial quench, as then $G_\text{F}(t)\equiv 1$. Nevertheless, these advantages come at a high price: Introducing the symmetric interference matrix
\begin{align}\label{eq:matrixA}
A_{ij}(t)=\frac{1}{Z}\sum_l \text{e}^{-\frac{\beta}{2} \epsilon_i^\text{i}} \langle \Psi_i | \Phi_l\rangle \, \text{e}^{-\text{i} \epsilon_l^\text{f} t} \langle \Phi_l | \Psi_j \rangle \, \text{e}^{-\frac{\beta}{2} \epsilon_j^\text{i}}\;,
\end{align}
where $\{|\Psi_i\rangle\}$ ($\{|\Phi_i\rangle\}$) are the eigenstates of the initial (final) Hamiltonian with eigenenergies $\epsilon^{\text{i}(\text{f}
)}_i$, allows us to represent the Loschmidt amplitudes \eqref{eq:LA_thermal} and \eqref{eq:LA_Fid} in a unified form.
While the interferometric Loschmidt amplitude $G_\text{I}(t)=\Tr\{A(t)\}$ is easily evaluated, its fidelity counterpart $G_\text{F}(t)=\Tr\{\sqrt{A(t)\cdot A(t)^*}\}$ is numerically far more expensive, because it requires calculating the square root of a large matrix at every time step.

\section{Quantum Loschmidt-echo return rate}\label{sec:rtilde}
Since on the one hand the interferometric definition $r_\text{I}(t)$ already shows cusps for a trivial quench where it probes the complex Lee-Yang zeros, and on the other hand computation of the fidelity return rate $r_\text{F}(t)$ entails an intractable numerical effort, we shall introduce a third finite-temperature return rate. While this choice will be specific to integrable systems like our infinitely connected model, it reconciles a simple physical motivation with a numerically efficient evaluation, and, furthermore, resolves the same phases as $r_\text{F}(t)$.
\subsection{Motivation and definition}
Let us consider the fully connected Ising model~\eqref{eq:classical}. Due to the fact that $[H,S^2]=0$, we conserve the total angular momentum independent of the initial and final values of $\Gamma_{{\text{i}},{\text{f}}}$. The ability to numerically treat system sizes of the order of several thousand sites relies directly on this fact. On the other hand, our quench protocol only allows for states within a fixed $S$-subspace to interfere during the time evolution. However, in the standard $r_\text{I}(t)$ we compute interferences of arbitrary $S$-subspaces:
\begin{equation}
r_\text{I}(t)=-\frac{2}{N}\ln\left|\sum_S G_S(t)\right|,
\end{equation}  
with 

\begin{align}
G_S(t)&=\Tr\left\{\text{e}^{-\text{i}H_{\text{f}}t}\Tr_{\cancel{S}}\rho_{\text{i}}\right\},
\end{align} 

\noindent the Loschmidt amplitude obtained for the subspace with total spin $S\in \{1/2,3/2,\ldots,N/2\}$ for odd $N$ without loss of generality. Here, $\Tr_{\cancel{S}}$ denotes a partial trace over all spin subspaces except that with spin length $S$. Obviously, in $r_\text{I}(t)$ all spin sectors interfere and can give rise to cusps despite the fact that the quench in $\Gamma$ cannot mix any states of different $S$.

Additionally, for a finite quench distance, $r_\text{I}(t)$ will show quite a rich behavior that is related only to the integrability of the fully connected model and is not expected to be found in a more generic system. If we expand $r_\text{I}(t)$ in a spectral representation
\begin{align}
r_\text{I}(t)=-\frac{2}{ N}\ln\left|\int \d\epsilon\; g(\epsilon) \text{e}^{-\text{i} \epsilon t}\right|,
\end{align}
we see that it is simply the Fourier transform of the modified density of states $g(\epsilon)$ that is given by 

\begin{align}
g(\epsilon)=\sum_S \frac{D(S)}{Z}\sum_{i,j} \left|\langle \Phi^S_j | \Psi^S_i\rangle\right|^2 \text{e}^{-\beta \epsilon_i^{\text{i},S}}\delta(\epsilon-\epsilon^{\text{f},S}_j)\;,
\end{align}
where again $\{|\Psi^S_i\rangle\}$ $\left( \{|\Phi_i^S\rangle\}\right)$ denote the eigenstates in the spin sector $S$ of the initial (final) Hamiltonian with energies $\epsilon_i^{{\text{i}}({\text{f}}),S}$. Like any ordinary density of states, $g(\epsilon)$ contains a superposition of Dirac distributions located at the actual final-Hamiltonian eigenvalues $\epsilon_i^{\text{f},S}$, but, importantly, here they carry weights proportional to the degeneracy factor $D(S)$ of the corresponding subspace. Due to its binomial behavior, see~\eqref{eq:degen}, $D(S)$ varies over several orders of magnitude between the different spin sectors. As such, we have to compute the Fourier transform of a very rough function that cannot become smooth even in the thermodynamic limit as the average level spacing remains of order one. In contrast, in a nonintegrable model, the huge degeneracies vanish and the typically exponentially small energy distances in the spectrum smoothen both $g$ and $r_\text{I}$. 
Thus, in order to investigate features that do not depend too crucially on the full permutation invariance of our model and to focus on cusps that are really evoked by our $S$-conserving quench protocol, we define a generalized Loschmidt echo $r_{\text{q}}(t)$ that sums all subspaces in phase
\begin{equation}\label{eq:rtilde}
r_{\text{q}}(t)=-\frac{2}{N}\ln\sum_S \left| G_S(t)\right|.
\end{equation} 
Quite importantly, this choice treats the mixing of the states by the quench and the resulting interferences on an equal footing. 
Furthermore, the sum here is always dominated by the subspace with the largest combination of degeneracy factor $D(S)$ times thermal weight of its ground state. This space can be found analytically in mean-field theory~\eqref{eq:s_max}.  As a consequence, thermal broadening disappears in the thermodynamic limit and all cusps in the Loschmidt echo become sharp signatures if they are for the system with $\Gamma^{\text{eff}}_{\text{i}(\text{f})}=\Gamma_{\text{i}(\text{f})}/\bar{s}$ for the quench at $T=0$, with $\bar{s}$ a solution of~\eqref{eq:s_max}.

Within the dominant subspace, like in every other subspace, all states have the same $D(S)$, cf. Sec.~\ref{sec:classical}, so the importance of a certain state during a quench depends only on its thermal weight factor and the overlaps with the eigenstates of the final Hamiltonian giving rise to a much smoother density of states and Fourier transform compared to the situation discussed in Sec.~\ref{sec:Loschmidt}. Finally, in a trivial quench, $r_{\text{q}}(t)$ will be a smooth function as the sum is now dominated by a single state which cannot give rise to interferences. This state is given by the ground state of the most important $S$-subspace.

\subsection{Comparison with fidelity return rate}
\begin{figure}[htp]
\includegraphics[width=\columnwidth]{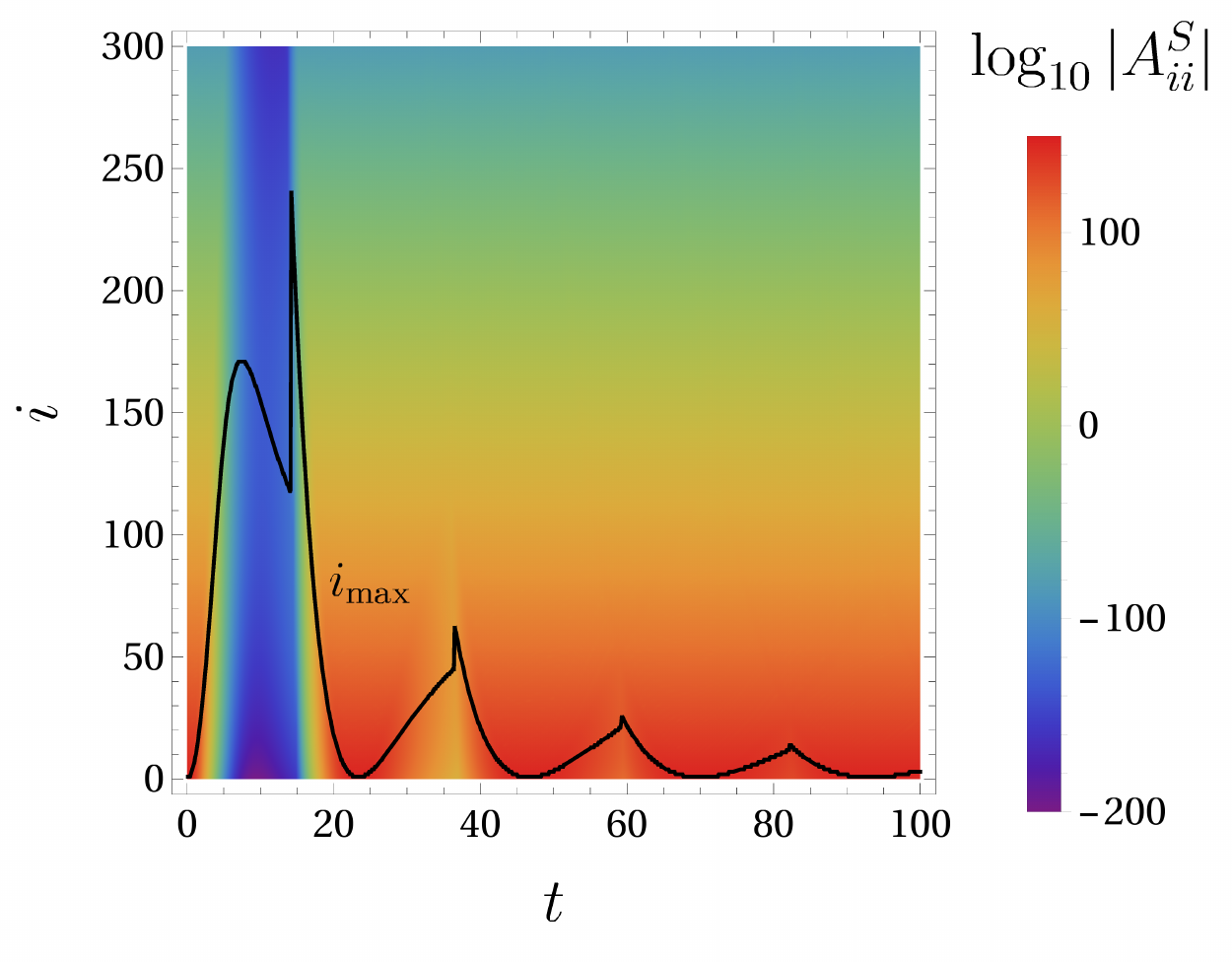}
\caption{(Color online) Decadic logarithm of the modulus of the diagonal elements $A^S_\text{ii}$ of the interference matrix \eqref{eq:matrixA} in the dominant spin subspace $S$, set by the classical saddle point \eqref{eq:s_max} as a function of time. Initially the largest component is given by $A^S_{11}$, however during the time evolution the index $i_\text{max}(t)$ is a nontrivial function that exhibits discontinuities, which coincide with the cusps in $r_\text{F}(t)$. Parameters for the quench are $\Gamma_\text{i}=0$ and $\Gamma_\text{f}=0.3$ at $\beta=5$ for a system of size $N=4000$.}
\label{fig:jump} 
\end{figure}
Taking into account that the total spin is conserved, the interference matrix $A$ from~\eqref{eq:matrixA} decouples into block diagonal form with a block $A^S$ for every spin sector. As a result both
\begin{align}
r_\text{q}(t)=-\frac{2}{N}\ln\sum_S\left|\Tr A^S\right|\;,
\end{align}
as well as
\begin{align}
r_\text{F}(t)=-\frac{2}{N}\ln\sum_S\Tr\sqrt{A^S\cdot {A^S}^*},
\end{align}
sum over all spin spaces in phase. Consequently, neither of these signals shows the large number of cusps $r_\text{I}(t)$ exhibits that cannot be associated with a physical phase. Due to the symmetric construction of $A^S$, its entries with largest moduli always appear on the diagonal.
Furthermore, without loss of generality, we sort the eigenstates $\{|\Psi^S_i\rangle\}$ in the fixed-$S$ subspace of $H_\text{i}$ by their energy such that the ground state is $|\Psi^S_1 \rangle$. At $t=0$, the matrix $A^S$ is diagonal, and due to the thermal weights, its maximal element is $A^S_{11}$.
At later times, the overlaps between different initial states remain suppressed, and, therefore, $A$ is still dominated by its diagonal elements. However, during the time evolution, interferences between the final states result in an oscillation of all elements $A^S_{ii}$ with a frequency that is given by the mean level spacing between those eigenstates of the final Hamiltonian with which the initial state $| \Psi_i^S \rangle$ has the largest overlaps. Since the spectrum of the final Hamiltonian is not perfectly linear, these oscillations are damped. For any $S$-subspace in the thermodynamic limit, we now define the function $i_\text{max}(t)\in(0,1]$ by requiring that $|A^S_{i_\text{max}(t)L,i_\text{max}(t)L}(t)|\geq|A^S_{ii}(t)|$ for all $i\in\mathbb{N}$ and $i\leq L=\text{dim}(A^S)$. Since $A^S$ is symmetric, the largest value of $M=A^S\cdot {A^S}^*$ will also be $M_{i_\text{max}L,i_\text{max}L}$. Consequently after a discontinuous change in $i_\text{max}$ the time-evolution of $A^S$ is governed by a new maximum, that in general evolves with a different slope, resulting in a discontinuity of the first derivative of the return function $r_\text{F}(t)$. Such a jump in the function $i_\text{max}(t)$ for a typical quench is presented in Fig.~\ref{fig:jump}. While $r_\text{q}(t)$ contains only the trace of $A^S$ and, therefore, also probes discontinuous changes in the diagonal elements of $A^S$, interferences between these can shift the cusp in time. However, after the first period of the return rate, the sum within the trace is once again evaluated almost completely in phase. Thus, before the end of the period, the corresponding cusp will also be visible in $r_\text{q}(t)$ and, consequently, $r_\text{F}(t)$ and $r_\text{q}(t)$ will always indicate the same dynamical phases. Larger quenches result in broader overlaps $\langle \Psi^S_i | \Phi^S_l\rangle$, whereas higher temperatures reduce the suppression of highly excited pre-quench eigenstates. Both effects lead to more destructive interferences in $A^S$. Based on this discussion, it becomes clear that larger quenches and higher temperatures give rise to earlier and more pronounced cusps in both return rates $r_\text{F}(t)$ and $r_\text{q}(t)$. Since the computation of the latter is several orders of magnitude faster, we will exclusively focus on the quantum return rate $r_\text{q}(t)$ for the remainder of this article.

\section{Results and discussion}\label{sec:results}

Using ED, we calculate the return rate~\eqref{eq:rtilde} and magnetization for several quenches of thermal initial states at various temperatures in order to construct the finite-temperature dynamical phase diagram shown in Fig.~\ref{fig:phasediagram} for the FC-TFIM.

\subsection{Quenches from the ferromagnetic phase}

\begin{figure*}[htp]
\centering
\hspace{-.25 cm}
\includegraphics[width=.335\textwidth]{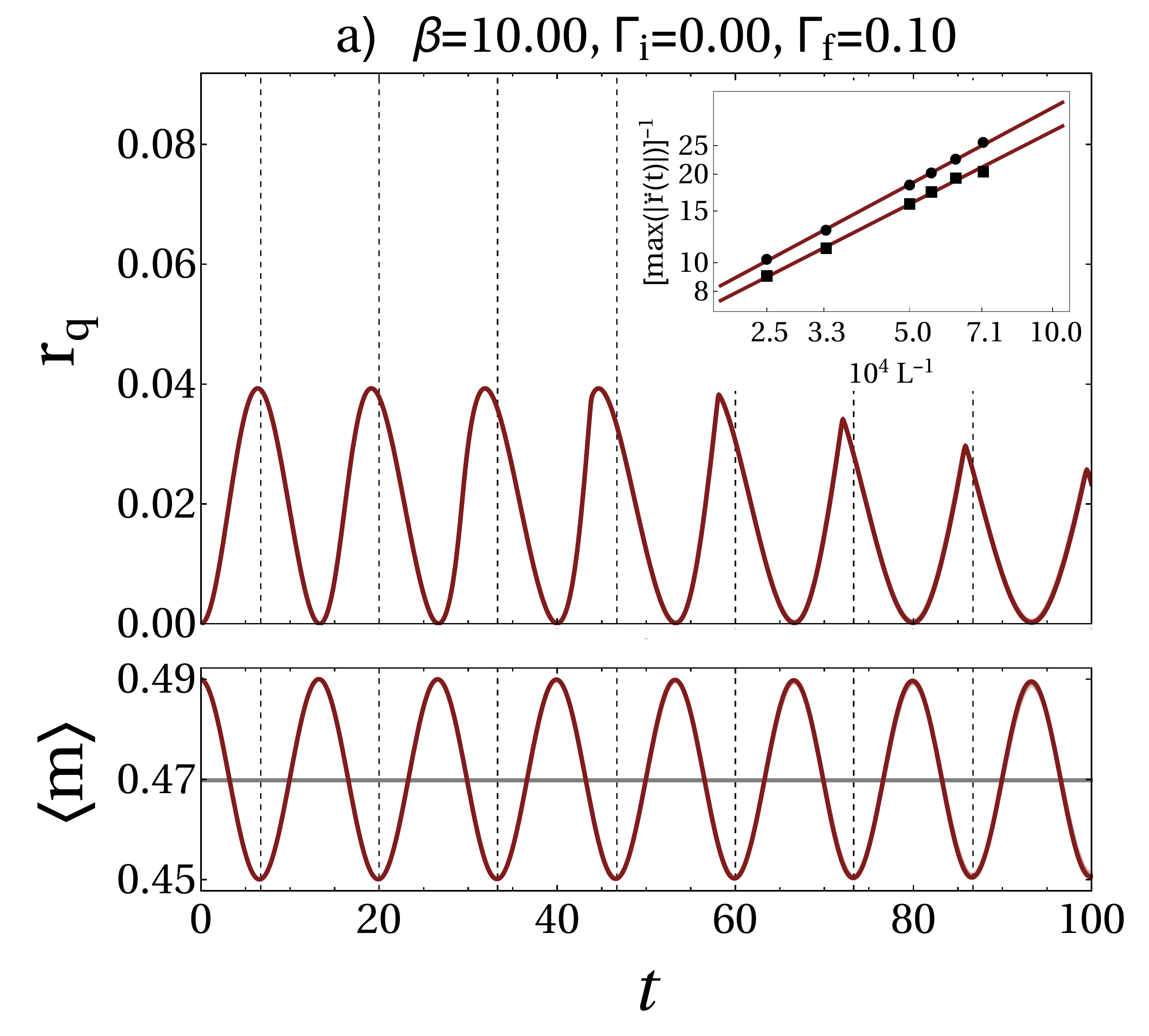}\quad
\hspace{-.4 cm}
\includegraphics[width=.335\textwidth]{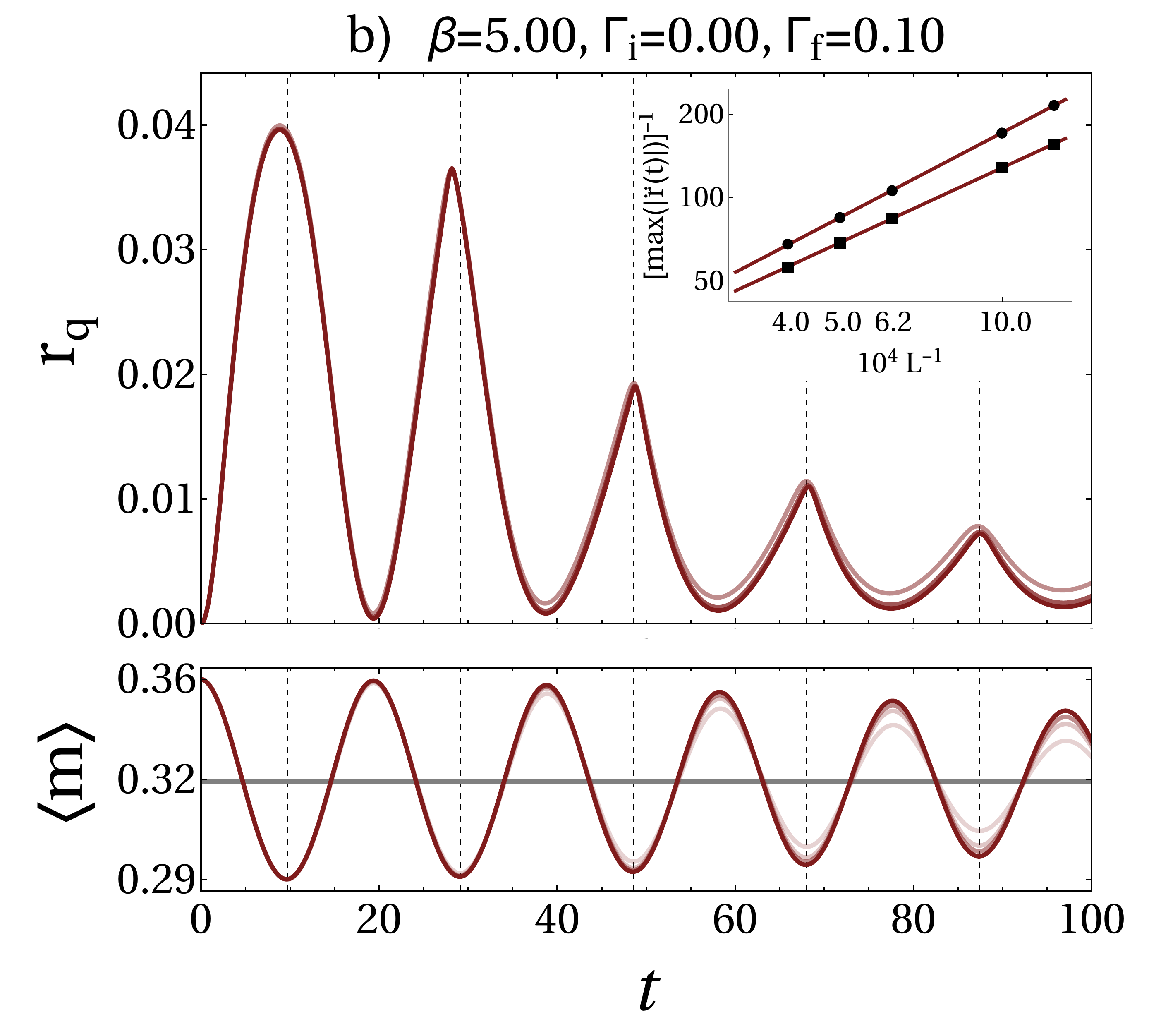}\quad
\hspace{-.4 cm}
\includegraphics[width=.335\textwidth]{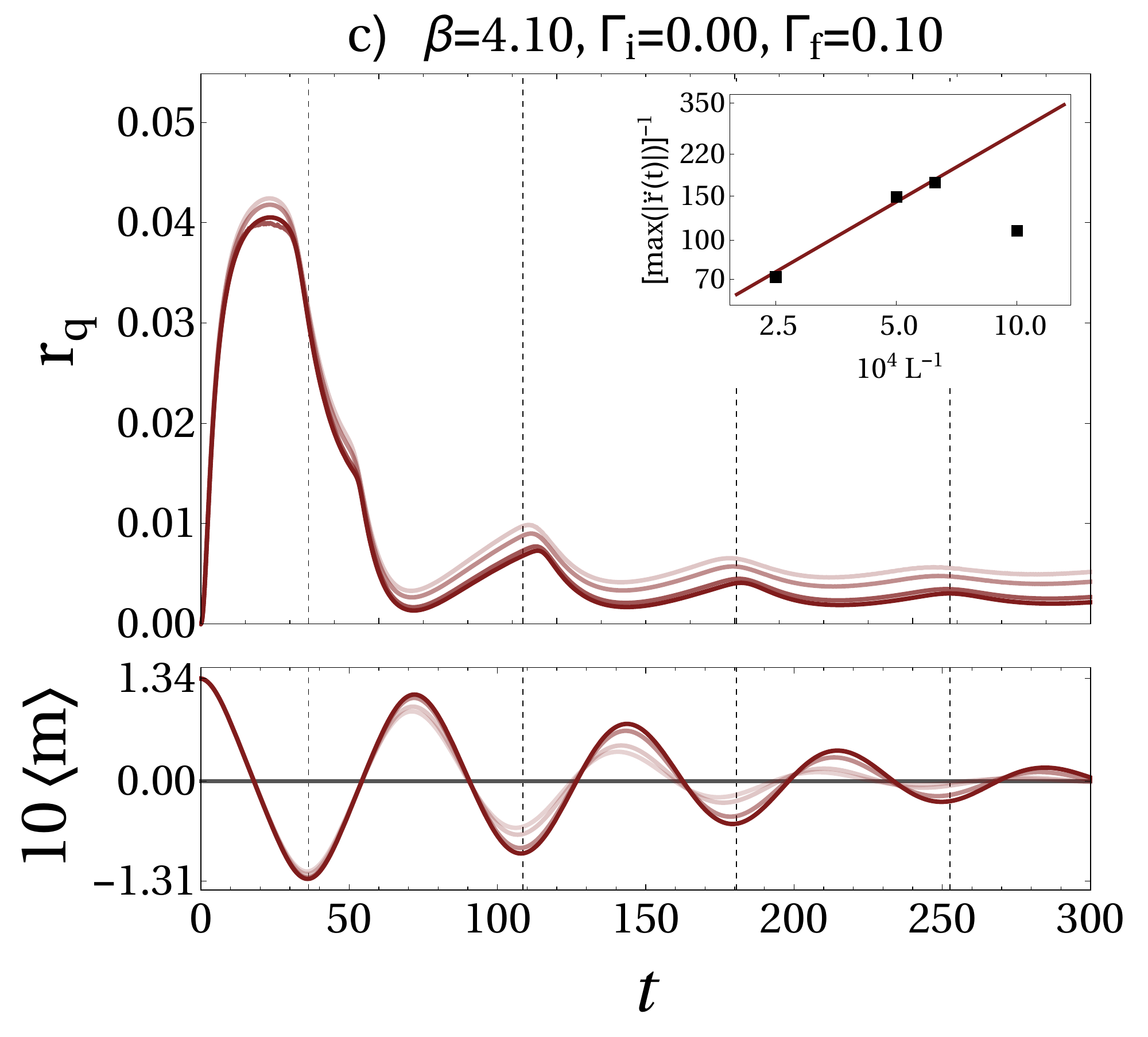}
\hspace{-.15 cm}
\caption{(Color online) Quantum quench in the FC-TFIM from $\Gamma_{\text{i}}=0$ to $\Gamma_{\text{f}}=0.1$ with initial ferromagnetic thermal state at inverse temperatures $\beta=10$, $5$, and $4.1$ (left to right) for various system sizes (light to dark red with increasing size) showing convergence. Even though at zero temperature this quench gives rise to an anomalous phase\cite{Homrighausen2017} in the FC-TFIM, as the temperature of the initial state is raised the anomalous phase transitions into its regular counterpart at temperatures above $T_{\text{c}}^{\text{d}}(\Gamma_{\text{i}}=0,\Gamma_{\text{f}}=0.1)$, cf.~\eqref{eq:T_c_dyn}. Corresponding magnetization plots show the agreement between the anomalous (regular) phase and the long-time ordered (disordered) Landau-type phase. The gray constant represents the time-averaged magnetization obtained from the classical equations of motion to which $\left\langle m(t) \right \rangle$ must converge in the long time limit. The grids connect the minima of the magnetization with the maxima of $r_\text{q}(t)$. Insets show the inverse curvature of each of the first two anomalous cusps in (a,b) and the first regular cusp in (c) in the return rate vanishing algebraically with system size, thereby indicating their sharpness, and thus true non-analyticity. For the sake of plot clarity, we only include the return-rate and magnetization plots for the four largest system sizes, where no artifacts due to limitations in the precision of the computation are visible: in (a) $N=1400,2000,3000,4000$, (b) $N=1000,1600,2000,2500$, and (c) $N=800,1000,1600,2000$.}
\label{fig:FM_quench_0.1} 
\end{figure*}

\begin{figure*}[htp]
\centering
\hspace{-.25 cm}
\includegraphics[width=.335\textwidth]{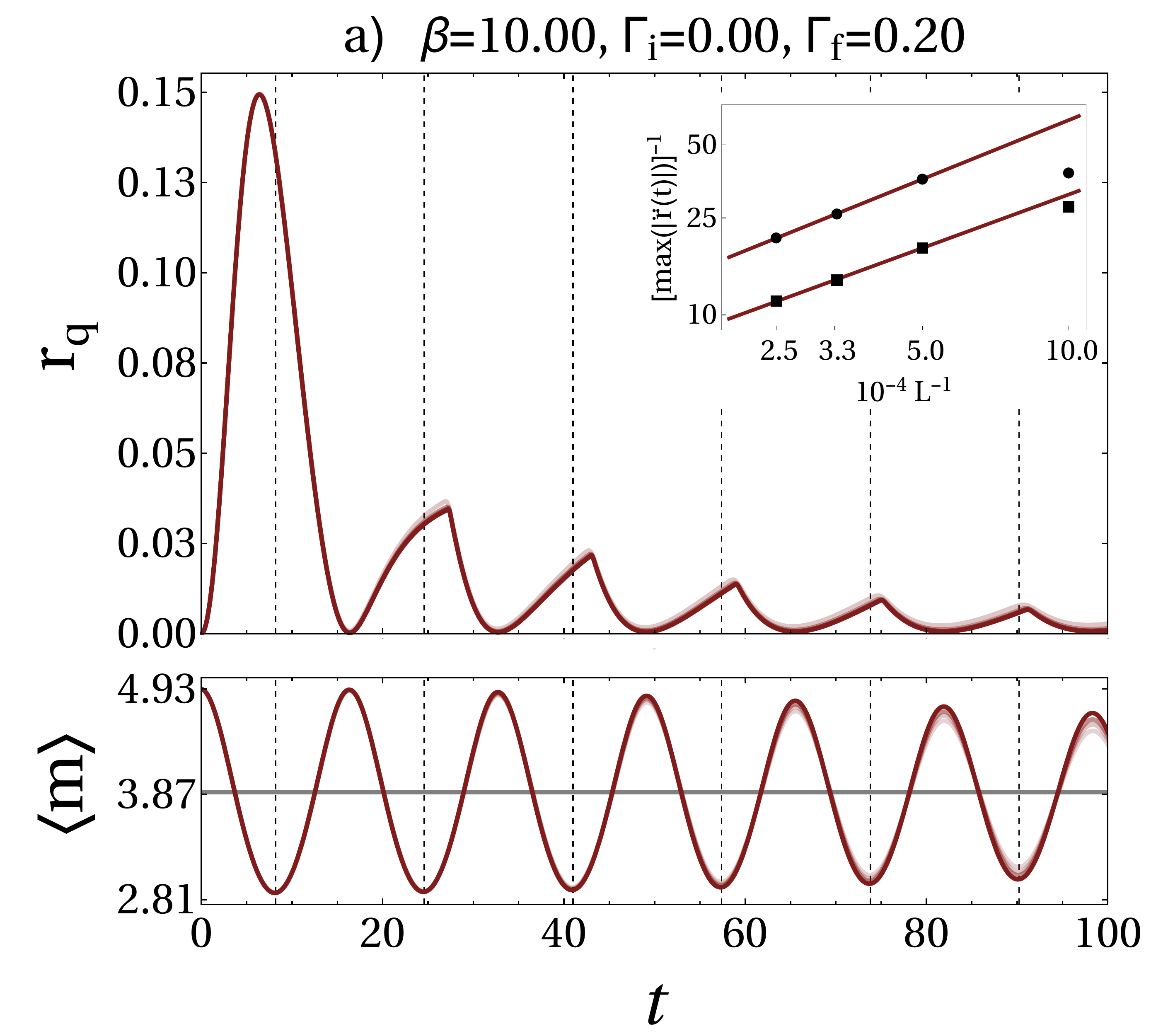}\quad
\hspace{-.4 cm}
\includegraphics[width=.335\textwidth]{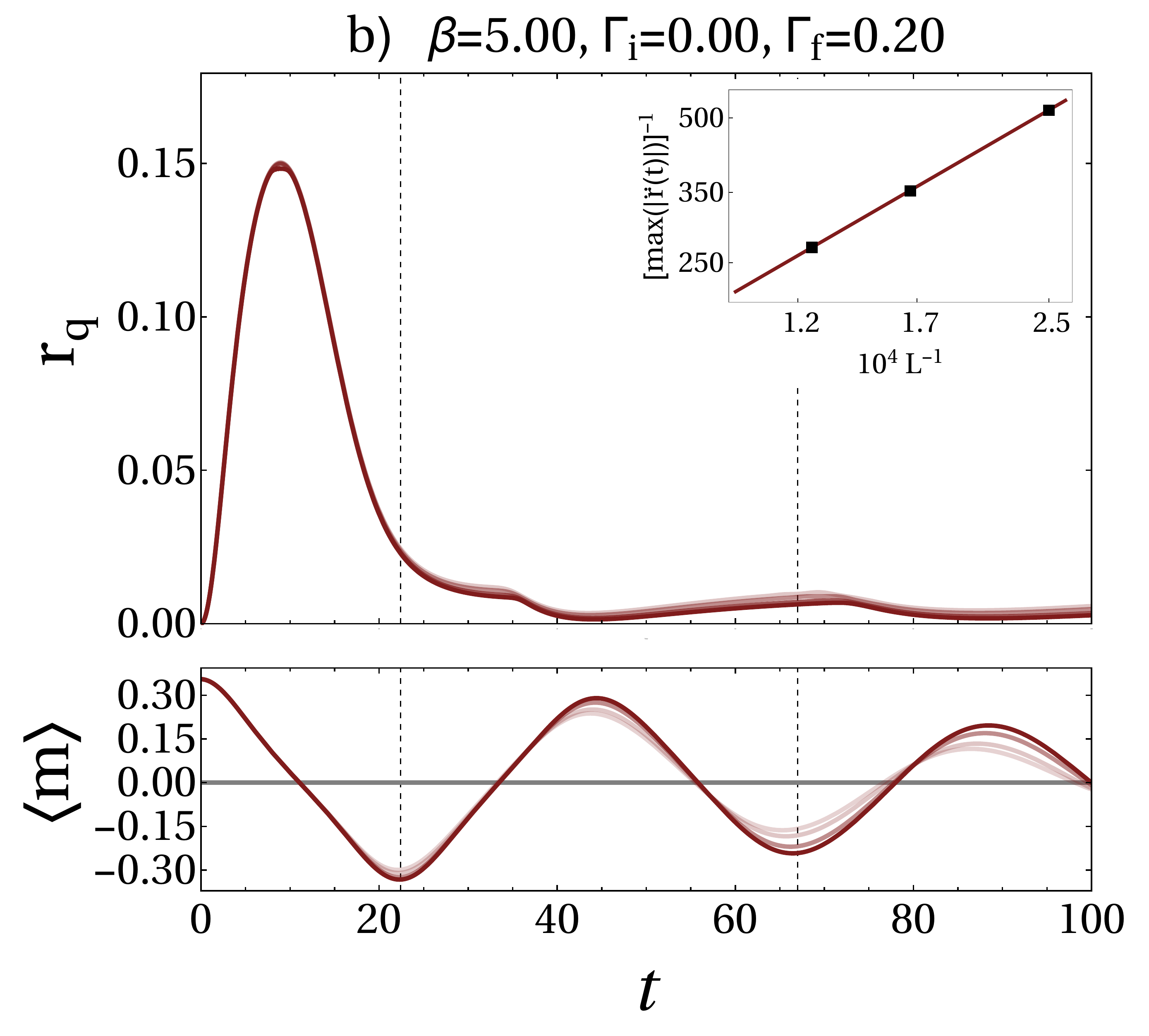}\quad
\hspace{-.4 cm}
\includegraphics[width=.335\textwidth]{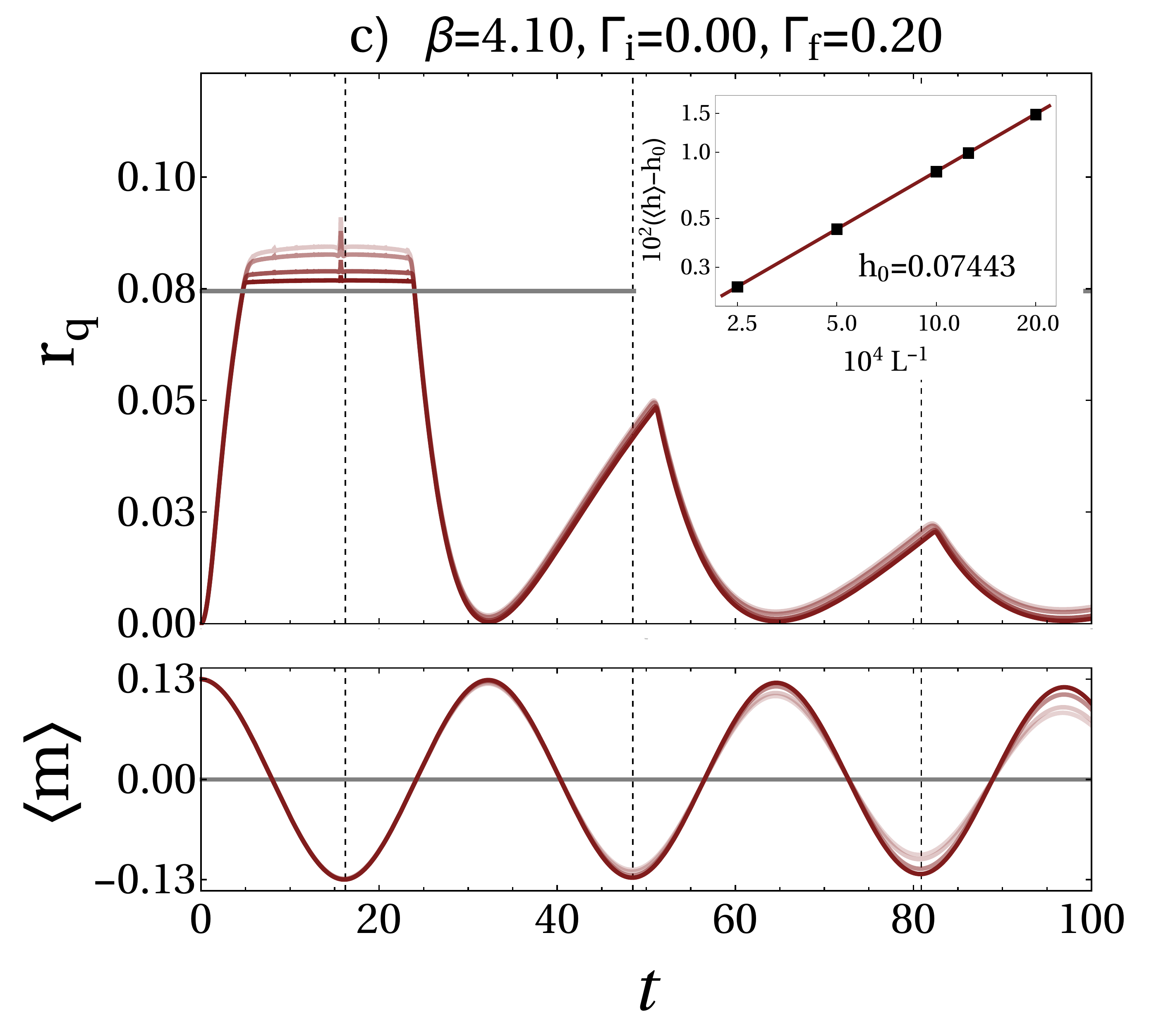}
\hspace{-.15 cm}
\caption{(Color online) Same as Fig.~\ref{fig:FM_quench_0.1} but with $\Gamma_\text{f}=0.2$. At low temperatures we again see an anomalous phase in (a), but now already at $T=0.2$ the phase is regular, which coincides with a zero infinite-time average of the magnetization. At the even higher temperature of $T=1/4.1$ where the return rate is even deeper in the regular phase, a thermal cutoff appears in the first peak occluding the cusp therein. Insets in (a) and (b) illustrate the divergence of the curvature of the first two anomalous cusps in (a) and the first regular cusp in (b), while the inset in panel (c) shows the algebraic convergence of the thermal cutoff height towards the analytical result for infinite system size as obtained from \eqref{eq:rmax}. System sizes are in (a) $N=1000, 2000, 3000, 4000$, in (b) $N=800, 1000, 1500, 2000$, and in (c) $N=800, 1000, 2000, 4000$.}
\label{fig:FM_quench_0.2} 
\end{figure*}

\begin{figure*}[htp]
\centering
\hspace{-.25 cm}
\includegraphics[width=.335\textwidth]{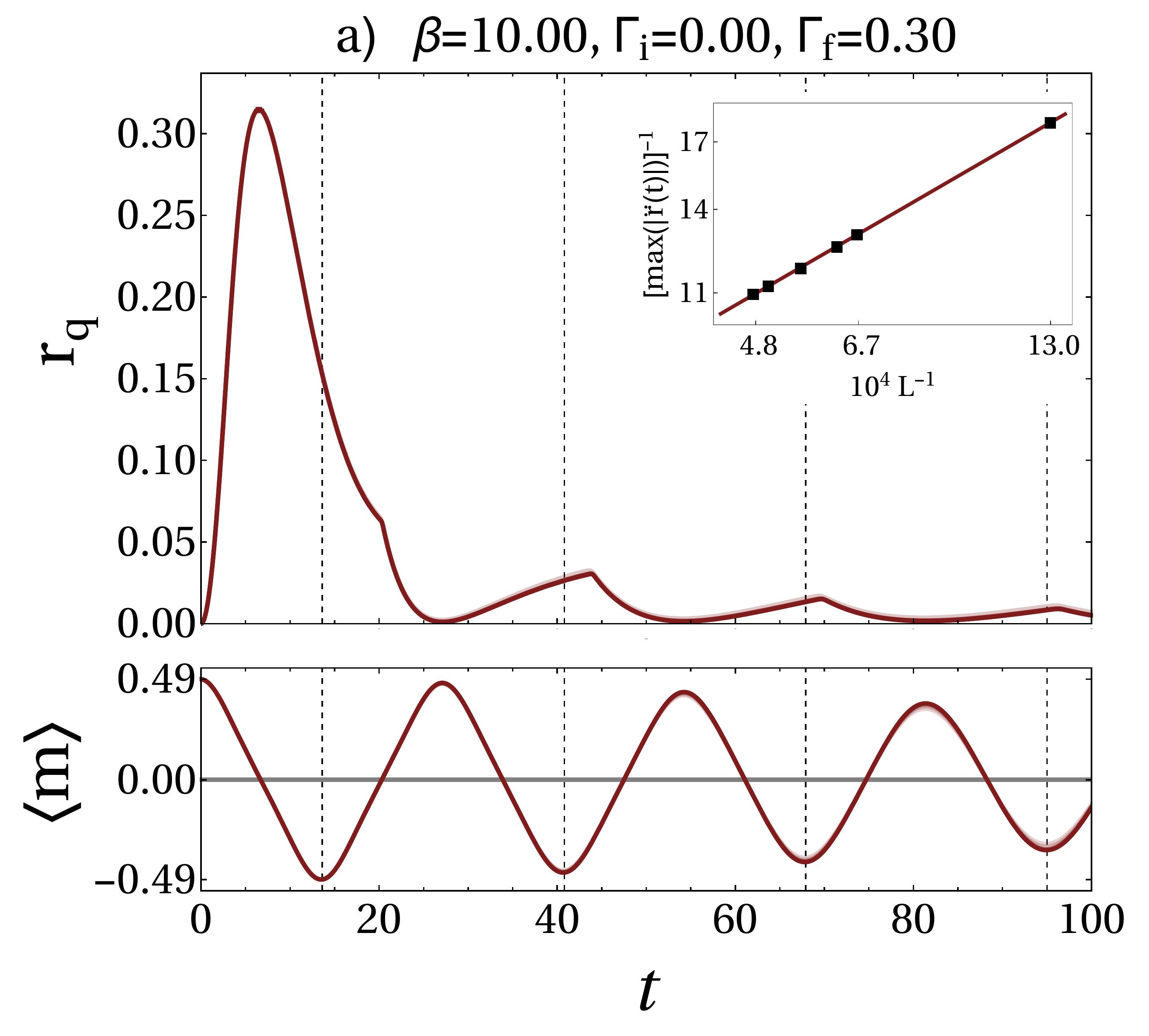}\quad
\hspace{-.4 cm}
\includegraphics[width=.335\textwidth]{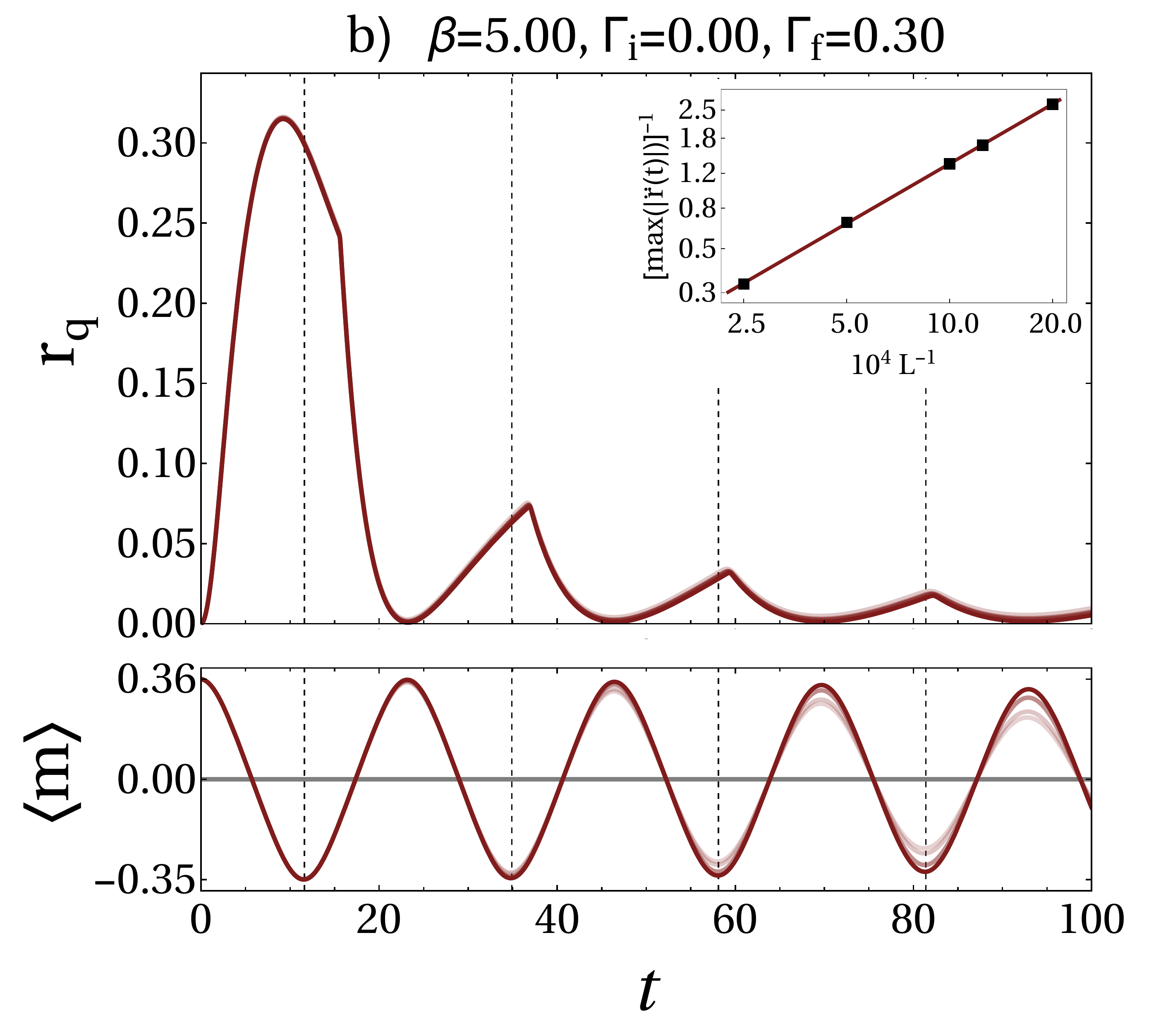}\quad
\hspace{-.4 cm}
\includegraphics[width=.335\textwidth]{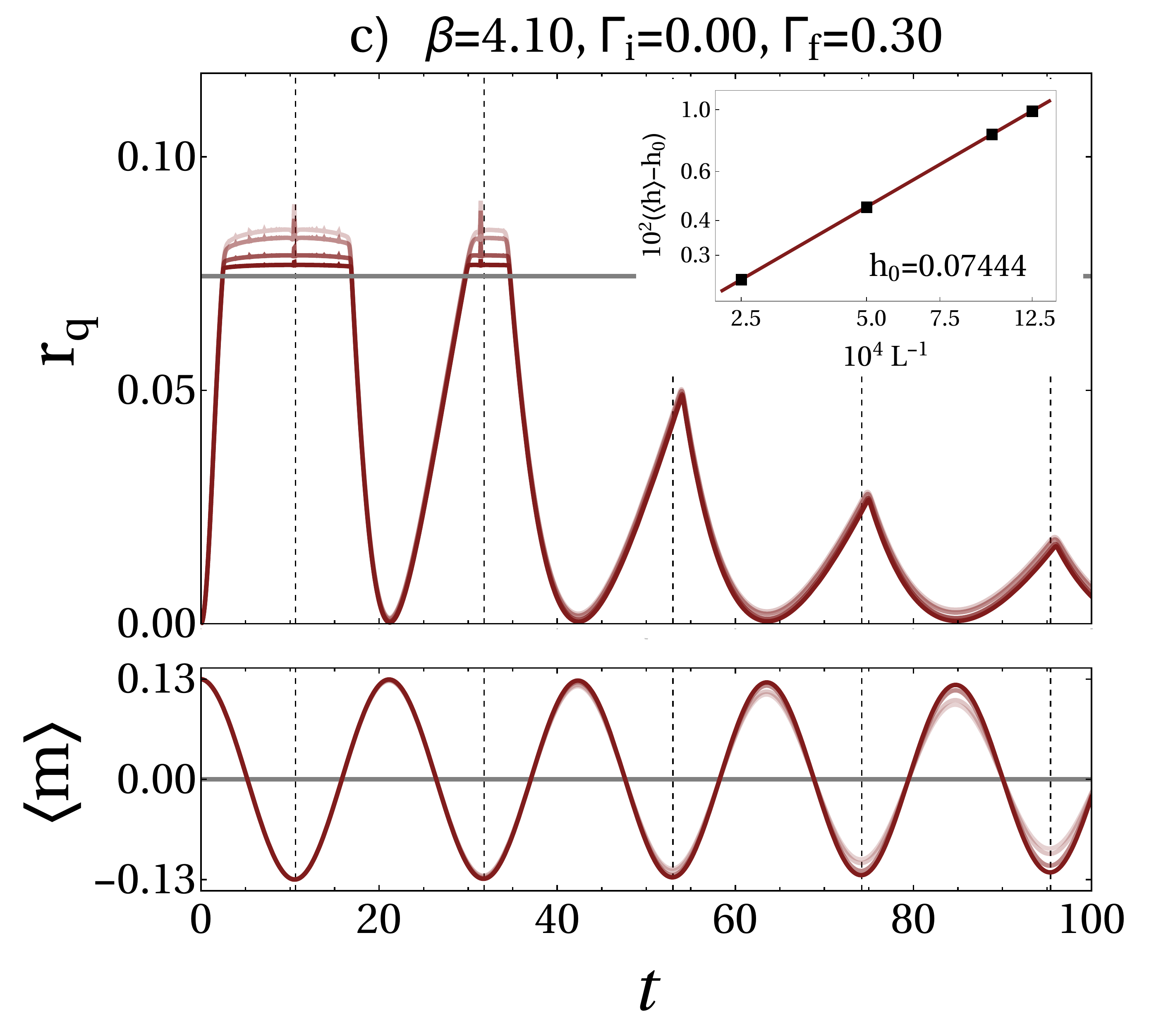}
\hspace{-.15 cm}
\caption{(Color online) Same as Fig.~\ref{fig:FM_quench_0.1} but with $\Gamma_\text{f}=0.3$. This quench gives rise to a regular phase even at $T=0$. At higher temperatures below $T_\text{c}^\text{e}|_{\Gamma\to0}=0.25$, the phase is therefore also regular. In (c) a thermal cutoff is visible in the first two peaks of the return rate. Insets are the same as in Fig.~\ref{fig:FM_quench_0.2}, but here only the curvature of the first cusp, that is relevant for the classification as a regular phase, is analyzed. The presented system sizes are $N=800, 1500, 1600, 1800$ in (a), $N=500, 800, 1000, 2000$ in (b), and $N=800, 1000, 2000, 4000$ in (c).}
\label{fig:FM_quench_0.3} 
\end{figure*}

We shall first present our results for quenches from a ferromagnetic thermal initial state, examples of which are shown in Figs.~\ref{fig:FM_quench_0.1},~\ref{fig:FM_quench_0.2}, and~\ref{fig:FM_quench_0.3}. At low temperatures and for short quench distances, the final state will still exhibit ferromagnetic order (see discussion in Sec.~\ref{sec:classical}). Following the quench, the initial magnetization vector, which for $\Gamma_\text{i}=0$ points along the positive $z$-direction with length $\bar{s}/2$ fixed by~\eqref{eq:s_max}, starts to precess within the upper hemisphere around a tilted mean magnetization. However, the equator will never be crossed, and, while dephasing will damp the precession, the mean magnetization $m$ cannot relax to zero. As our numerical investigation shows, this behavior is always accompanied by an anomalous phase in the return rate, where cusps appear only after its first minimum at finite time. The anomalous phase has previously been reported on in the FC-TFIM and one-dimensional transverse-field Ising model with power-law interactions for quenches starting from a ferromagnetic ground state in the case of an ordered final steady state.\cite{Homrighausen2017,Halimeh2016b} For the short quench distance $\Gamma_\text{i}=0\to\Gamma_\text{f}=0.1$ and the low temperature $T=0.1$ in Fig.~\ref{fig:FM_quench_0.1}(a), the return rate shows a strongly anomalous behavior characterized by many smooth periods before the appearance of the first cusp. The inset demonstrates the finite-size scaling of the curvature of the first two cusps, which is clearly consistent with the algebraically divergent model $\propto L^\alpha$ with $\alpha>0$ used in the fit. Preparing our initial state at $T=0.2$, on the other hand, we see in Fig.~\ref{fig:FM_quench_0.1}(b) that the same quench leads to a return rate where only the first peak is smooth, and thereafter every period of the return rate contains one cusp. This indicates that the higher the preparation temperature, the \textit{closer} we are to a regular phase. Indeed, upon further increasing the preparation temperature to $T=1/4.1$, which is very close to the equilibrium thermal critical point $T_\text{c}^\text{e}|_{\Gamma\to0}=1/4$, the anomalous phase disappears and is replaced by its regular counterpart, as shown in Fig.~\ref{fig:FM_quench_0.1}(c). At the same time the ferromagnetic order, which is already compromised by thermal fluctuations in the initial state, is lost completely in the final state. A closer investigation of the behavior for temperatures between $T=0.2$ and $T=1/4.1$ shows that, within our numerical precision, DPT-I and DPT-II coincide perfectly.

For the larger quench distance $\Gamma_\text{i}=0\to\Gamma_\text{f}=0.2$ and at small temperatures, we observe an anomalous phase as shown in Fig.~\ref{fig:FM_quench_0.2}(a). However, in accordance with the DPT-I, the regular behavior of the return rate with cusps in every peak (see Fig.~\ref{fig:FM_quench_0.2}(b)) appears at smaller temperatures than in the smaller quench of Fig.~\ref{fig:FM_quench_0.1}. At even higher temperatures, but still below $T_\text{c}^\text{e}|_{\Gamma\to0}$, something unexpected happens in the return rate: whereas the quench ends up in a state that is deep within the paramagnetic phase where one expects a regular behavior of the return rate, a chipped-off first peak is observed. This can be explained by noting that for high preparation temperatures in the ferromagnetic phase the dominant subspace becomes very short. This in turn implies that the contribution of the short $S$-subspaces with $S\sim \mathscr{O}(1)$ instead of $S\sim \mathscr{O}(N)$ can become large enough to be resolved in the return rate. However, within these subspaces, no contributions to the Loschmidt echo that are exponentially small in system size can be generated due to the absence of enough interfering terms in the sum. Instead, these manifest as sharp (logarithmically divergent) signatures on top of the return rate that vanish $\propto1/N$. As a result, the return rate is limited at (almost) all times by a maximal value $r_\text{q}^\text{max}$. By an argument similar to the one employed in Sec.~\ref{sec:Loschmidt}, one obtains for this value in the thermodynamic limit
\begin{align}\label{eq:rmax}
r_\text{q}^\text{max}=\beta\sqrt{m^2+\Gamma_\text{i}^2}+2\ln{\left(\frac{1+\text{e}^{-\beta\sqrt{m^2+\Gamma_\text{i}^2}}}{2}\right)}\;.
\end{align}
This expression can be confirmed numerically by the finite-size scaling in the insets of Figs.~\ref{fig:FM_quench_0.2}(c) and~\ref{fig:FM_quench_0.3}(c). These are performed by fitting a constant $h_0$ plus an algebraic
decay to the average height of the plateau on top of the first peak.
The obtained values for $h_0$ agree very well with $r_\text{q}^\text{max}
\approx 0.07445$.
For quenches within the anomalous phase, the classical magnetization vector never crosses the equator of the Bloch sphere and our numerical simulations show that the return rate never grows to a value sufficiently large so as to resolve $r_\text{q}^\text{max}$. Consequently, the thermal cutoff can only be seen for quenches from a ferromagnetic to a paramagnetic state, \textit{i.e.}~only in the regular phase. At the same time not every quench will be affected by $r_\text{q}^\text{max}$, but rather predominantly those involving large quench distances where the overlap between initial and final state is generally smaller, as can be witnessed in Figs.~\ref{fig:FM_quench_0.1},~\ref{fig:FM_quench_0.2}, and~\ref{fig:FM_quench_0.3}, where the latter shows the large quench from $\Gamma_\text{i}=0\to\Gamma_\text{f}=0.3$. Despite the cutoff, the underlying phase is still regular in both of Figs.~\ref{fig:FM_quench_0.2}(c) and~\ref{fig:FM_quench_0.3}(c), as can be seen by decreasing the preparation temperature. For lower temperatures in the regular phase, the cusp is located on the shoulder of the first maximum. Upon varying the temperature, it moves up the trailing slope until it reaches the simultaneously decreasing value of $r_\text{q}^\text{max}$. From this point onwards it will be hidden by the thermal cutoff. For an illustrative example see Appendix~\ref{app:illustration}. At the same time the order-parameter average is zero at infinite time indicating the equivalence of DPT-I and DPT-II. Note that the quench of Fig.~\ref{fig:FM_quench_0.3} already at zero temperature gives rise to the regular phase,\cite{Homrighausen2017} and, as such, all temperatures $T<1/4$ result in a regular phase.

As has been established in previous analytical\cite{Heyl2013} and numerical\cite{Zunkovic2016b,Halimeh2016b,Halimeh2017a,Homrighausen2017} studies, the periodicity of the return rate coincides with that of the magnetization, whether the underlying phase is anomalous ($\equiv$ ordered) or regular ($\equiv$ disordered), as can be seen in the bottom in Figs.~\ref{fig:FM_quench_0.1},~\ref{fig:FM_quench_0.2}, and~\ref{fig:FM_quench_0.3}.
Physically, the smallest overlap between final and initial states is obtained whenever the classical magnetization vector is furthest from its original orientation. This happens for times $t=\omega_\text{mag}^{-1}n\pi/2 $, where $n$ is an odd positive integer and $\omega_\text{mag}$ is the frequency of the precession of the magnetization.

Beyond the representative examples shown here, our extensive numerical simulations indicate that the DPT-I and DPT-II dynamical critical lines, to high precision, fully overlap in the $T-\Gamma_\text{f}$ phase diagram for $\Gamma_\text{i}<\Gamma_\text{c}^\text{e}(T)$.
Moreover, this dynamical critical line can be directly connected to the equilibrium critical line. In fact, it exactly coincides with the equilibrium critical line if, at a temperature $T<T_\text{c}^\text{e}|_{\Gamma\to0}$, the ferromagnetic thermal state is prepared at $\Gamma_\text{i}=\Gamma_\text{c}^\text{e}(T)-\delta$, with $\delta\to0^+$.

\begin{figure*}[htp]
\centering
\hspace{-.25 cm}
\includegraphics[width=.335\textwidth]{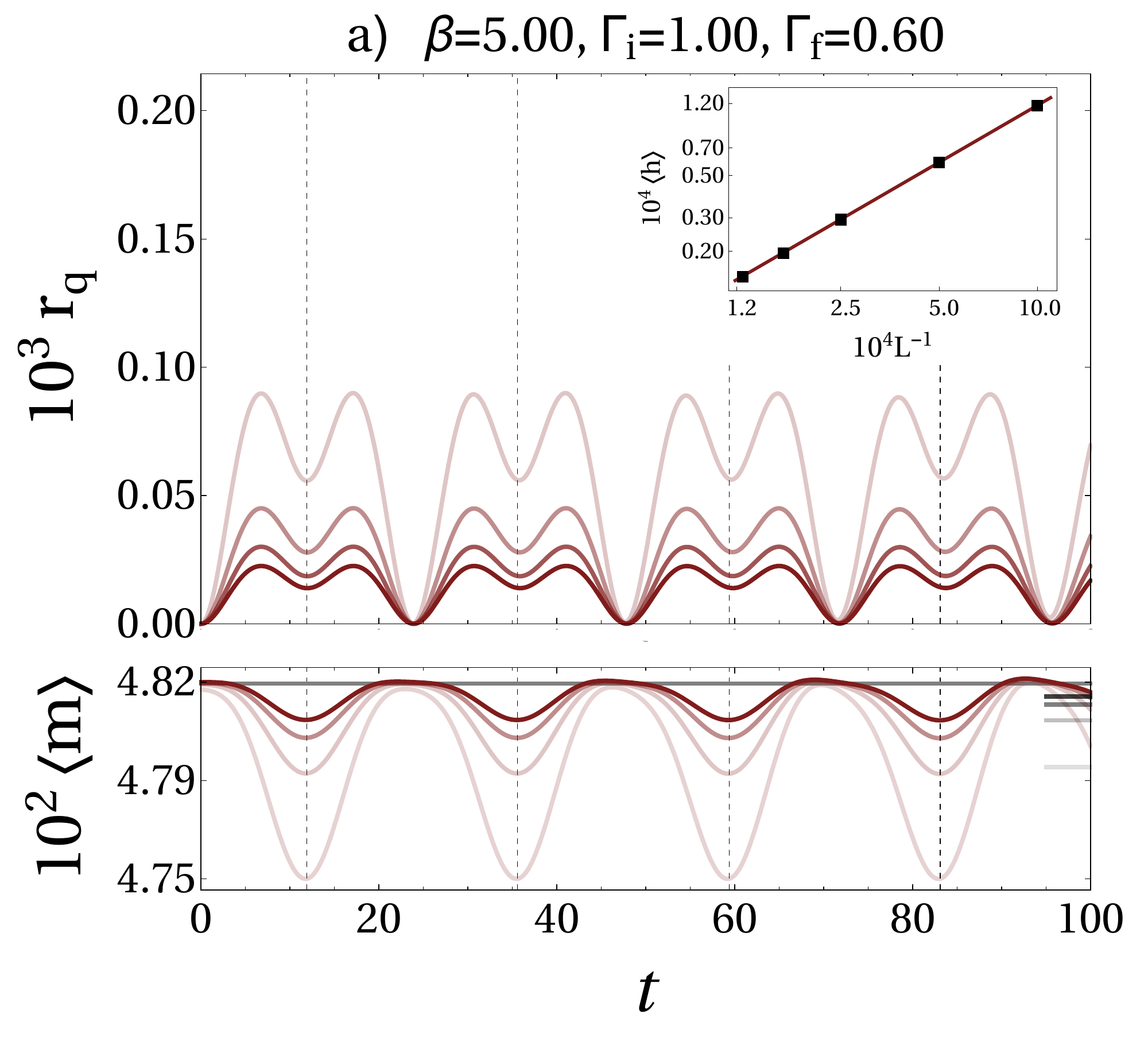}\quad
\hspace{-.4 cm}
\includegraphics[width=.335\textwidth]{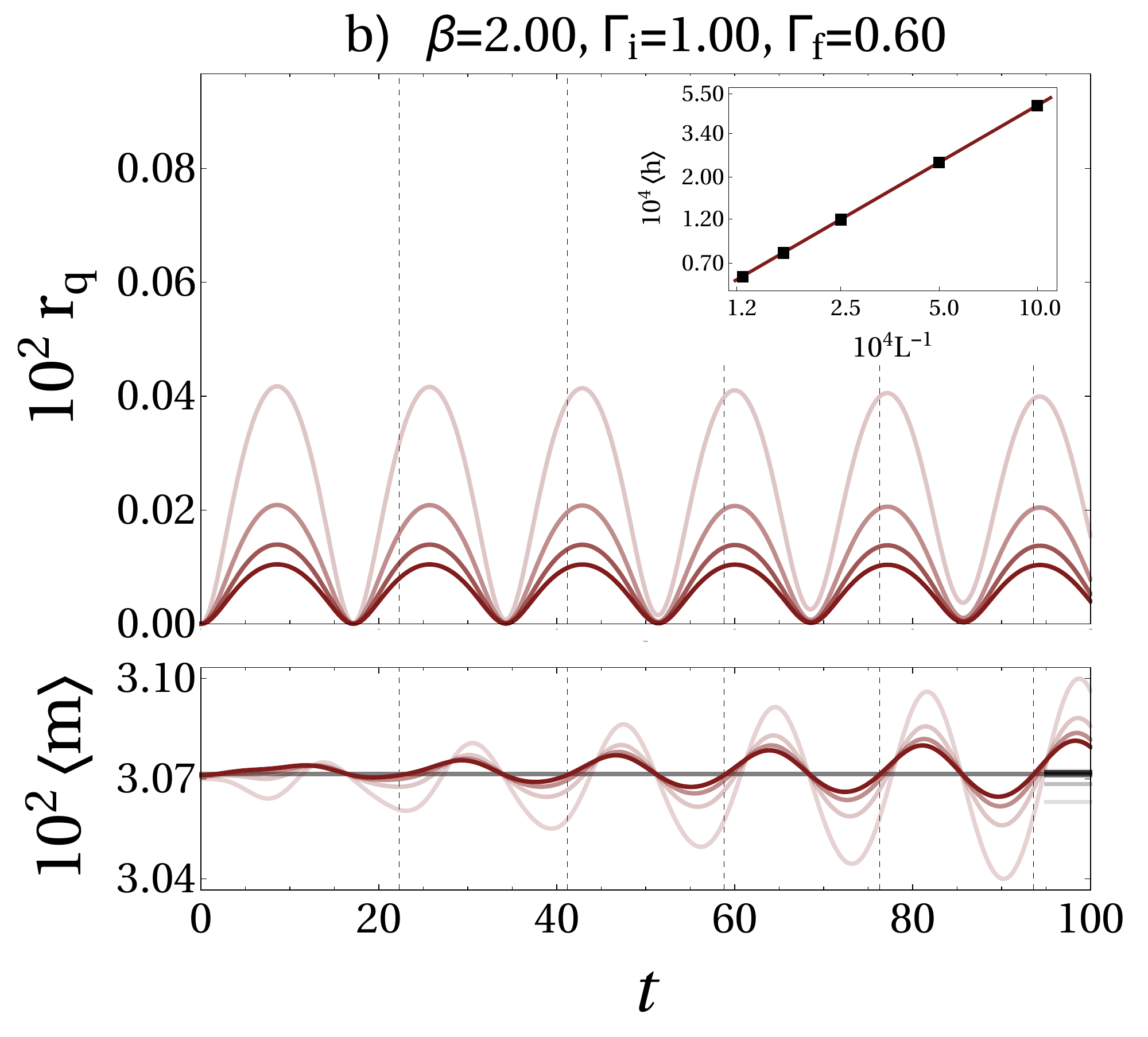}\quad
\hspace{-.4 cm}
\includegraphics[width=.335\textwidth]{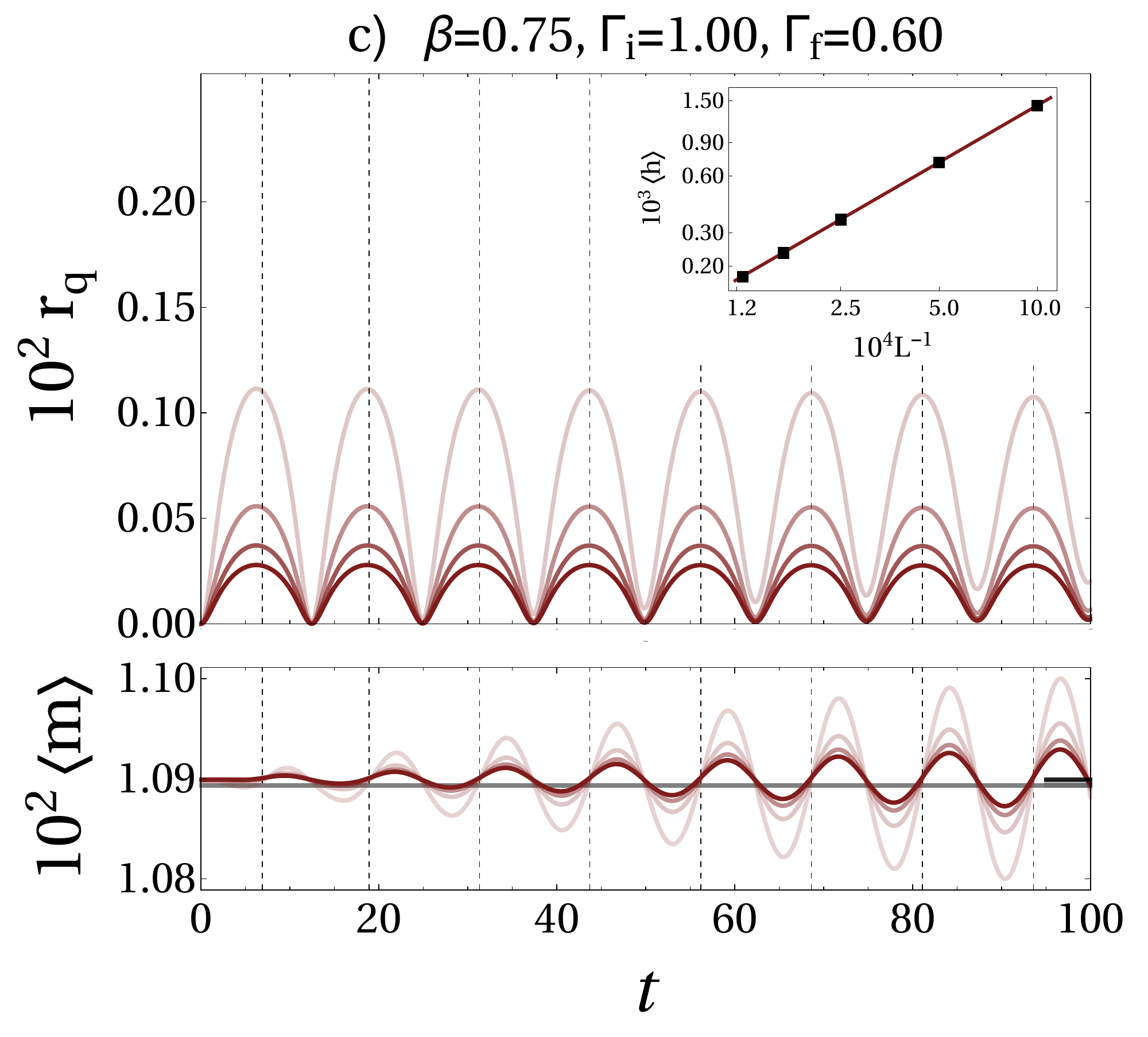}
\hspace{-.15 cm}
\caption{(Color online) Quantum quench in the FC-TFIM from $\Gamma_{\text{i}}=1$ to $\Gamma_{\text{f}}=0.6$ at inverse temperatures $\beta=5$, $2$, and $0.1$ for $N= 2000, 4000, 6000, 8000$. This quench is within the paramagnetic phase at any temperature, and thus the return rate exhibits the trivial phase which scales to zero in the thermodynamic limit. The insets show the average amplitude of $r_\text{q}(t)$ over the first period as a function of system size, showing clear algebraic decay. The infinite-time magnetization of an infinite system is constant at the seeding value, indicating a disordered infinite-time steady state (see Sec.~\ref{sec:results}). The convergence of the infinite-time magnetization with increasing system size towards this value is indicated with increasingly opaque gray lines at the right edge of the magnetization plot.}
\label{fig:PM_quench_0.6} 
\end{figure*}

\begin{figure*}[htp]
\centering
\hspace{-.25 cm}
\includegraphics[width=.335\textwidth]{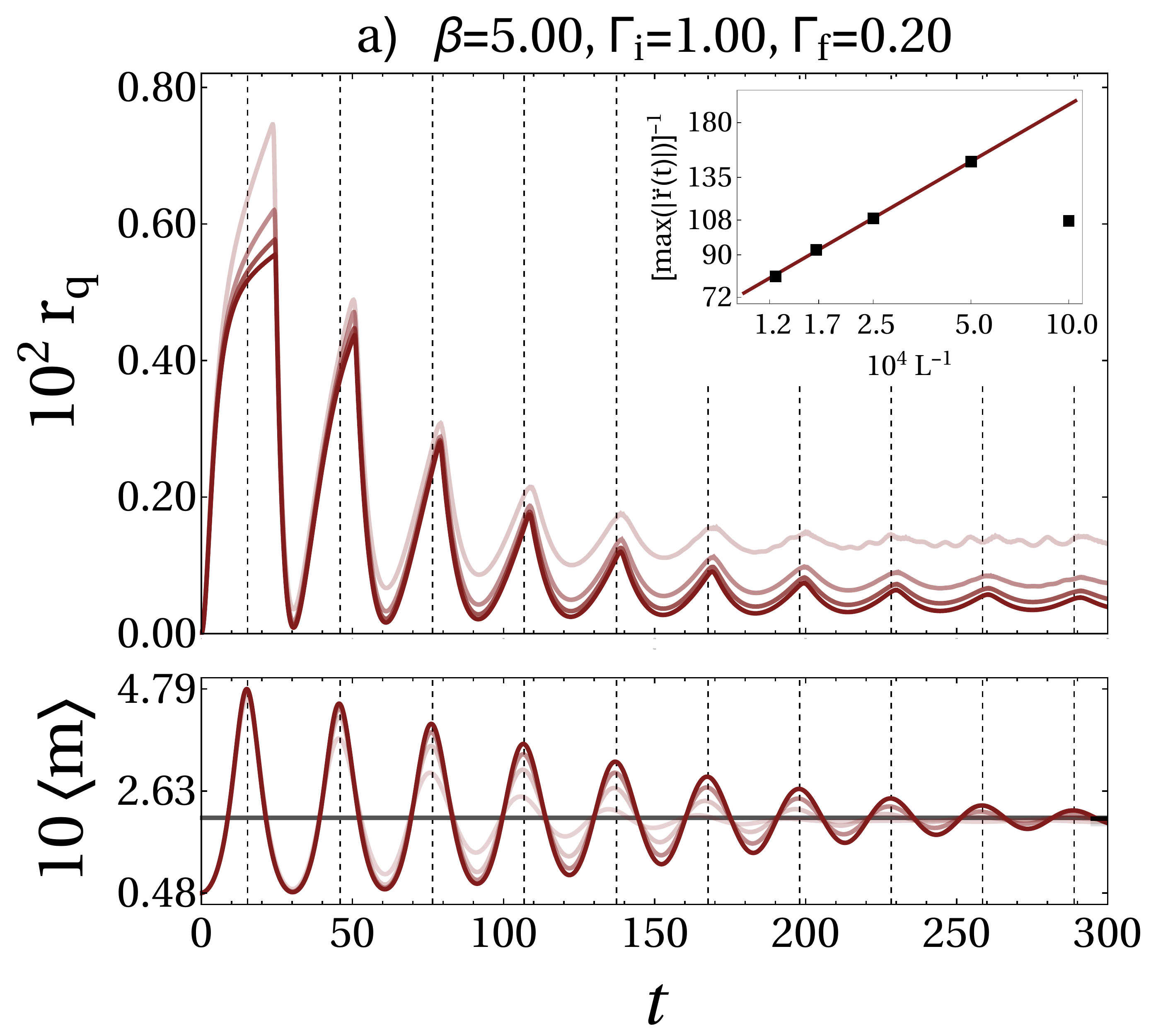}\quad
\hspace{-.4 cm}
\includegraphics[width=.335\textwidth]{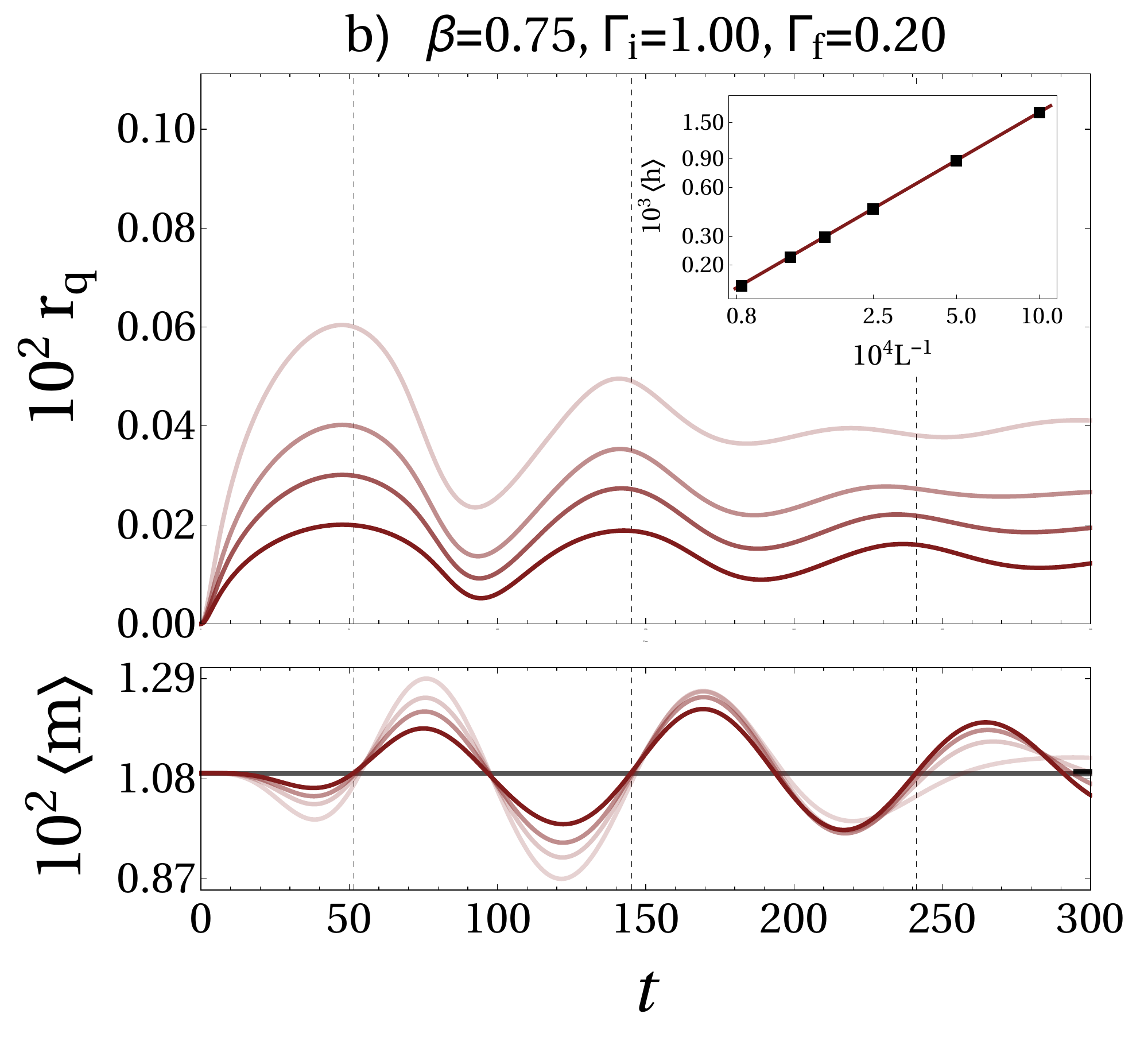}\quad
\hspace{-.4 cm}
\includegraphics[width=.335\textwidth]{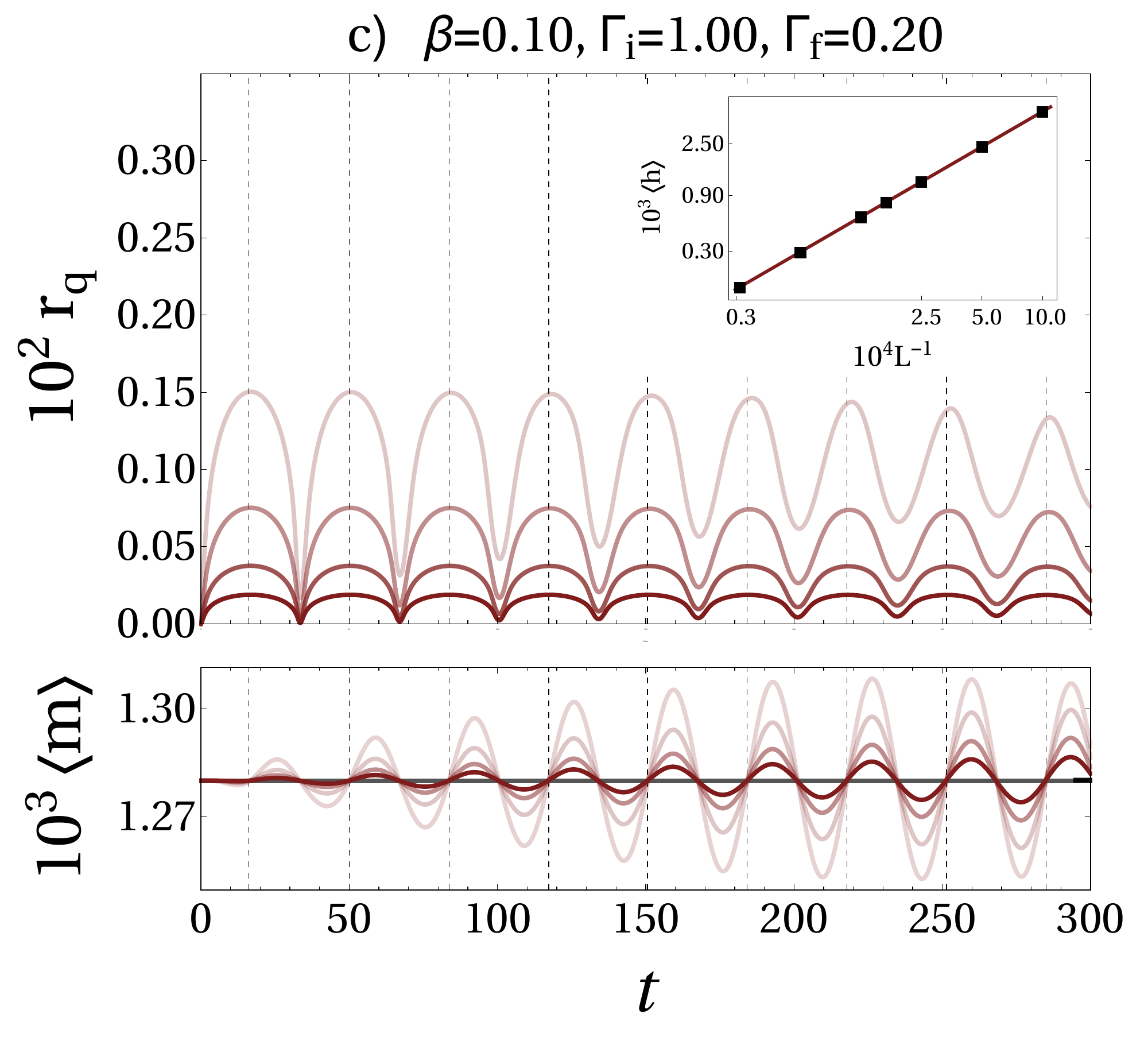}
\hspace{-.15 cm}
\caption{(Color online) Same as Fig.~\ref{fig:PM_quench_0.6} but for $\Gamma_\text{f}=0.2$. At the temperature $T=0.2$, a regular phase emerges in (a) coinciding with an ordered infinite-time steady state. The inset in (a) shows the finite-size scaling of the curvature of the first cusp, indicating its algebraic divergence with system size, and, therefore, the true non-analyticity of the cusp in the thermodynamic limit. The depicted system sizes are: (a) $N=2000, 4000, 6000, 8000$, (b) $N=4000, 6000, 8000, 12000$, (c) $N= 4000, 8000,16000, 32000$.}
\label{fig:PM_quench_0.2} 
\end{figure*}

\begin{figure*}[htp]
\centering
\hspace{-.25 cm}
\includegraphics[width=.335\textwidth]{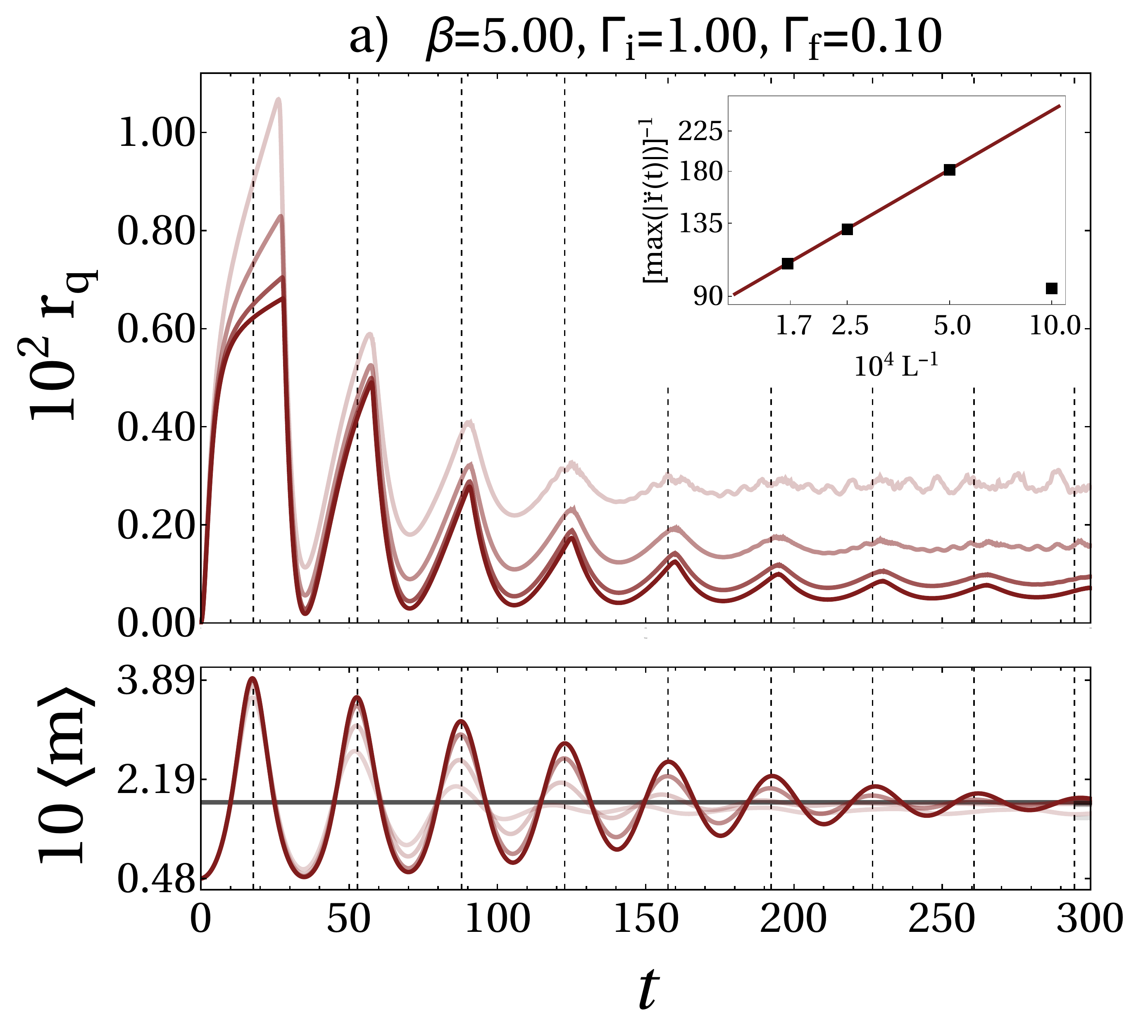}\quad
\hspace{-.4 cm}
\includegraphics[width=.335\textwidth]{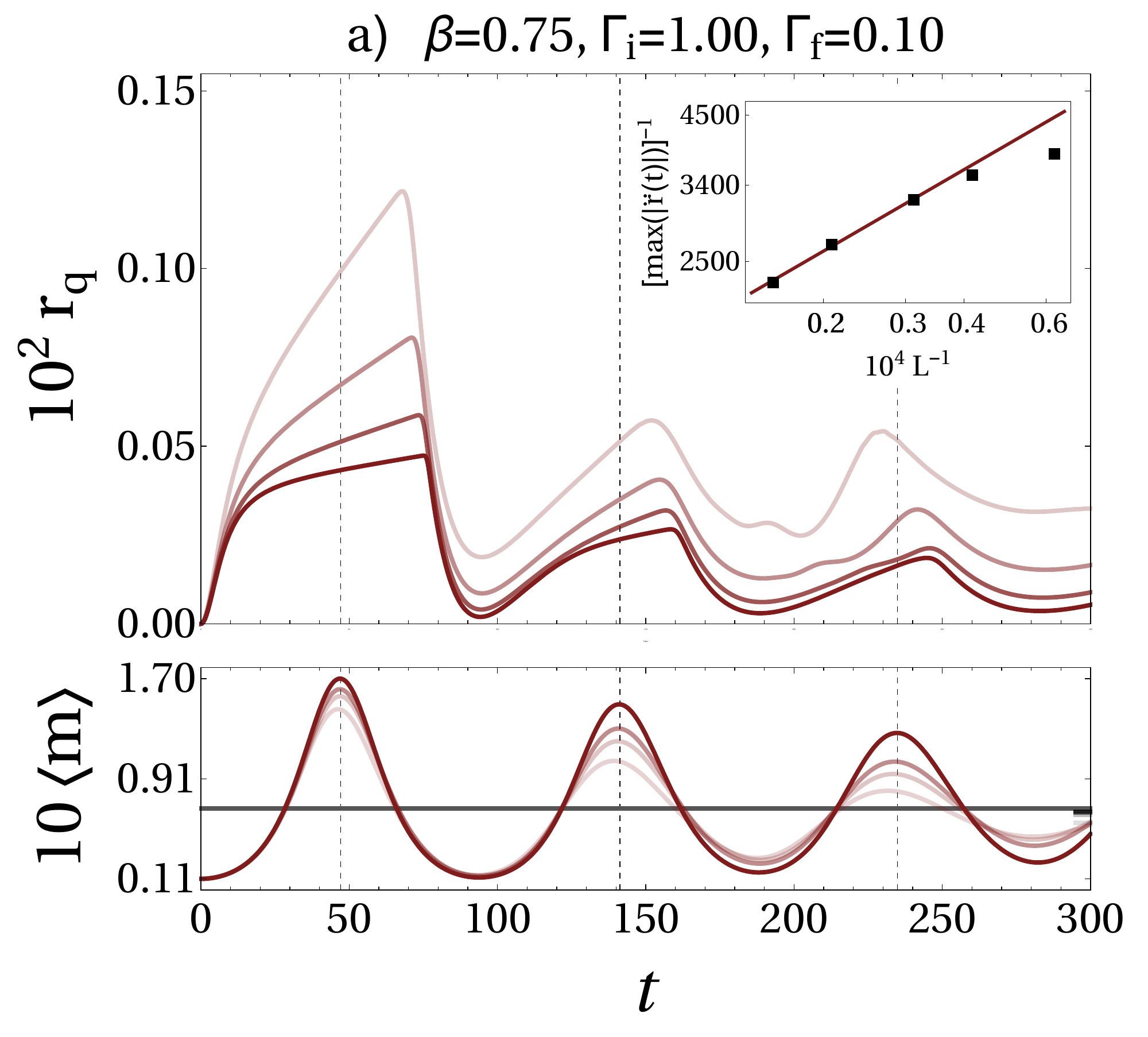}\quad
\hspace{-.4 cm}
\includegraphics[width=.335\textwidth]{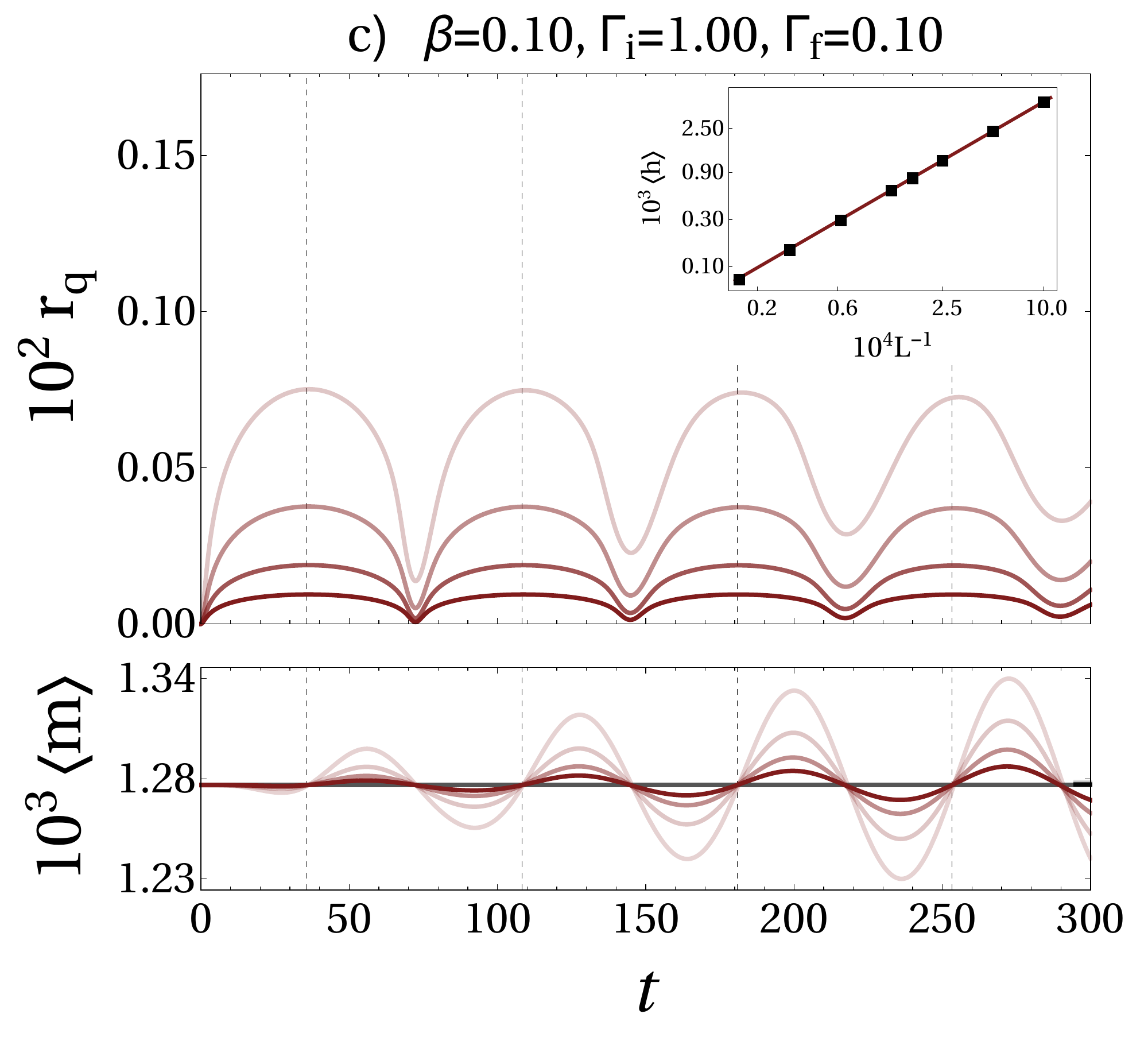}
\hspace{-.15 cm}
\caption{(Color online) Same as Fig.~\ref{fig:PM_quench_0.6} but with $\Gamma_\text{f}=0.1$. Even though in equilibrium $T=4/3$ corresponds to a paramagnetic state at any value of the transverse field, for this quench it is already low enough to give rise to a regular phase in the return rate, which coincides with an ordered infinite-time steady state. The insets in panels (a) and (b) show the finite-size scaling of the curvature of the first cusp, which diverges algebraically with system size, while the inset in (c) depicts how the average height of the first peak decays to zero algebraically as the system size is increased. The plots show system sizes of $N= 1000, 2000, 4000, 6000$ in (a), as well as $N=8000, 16000 ,32000, 64000$ in (b,c).}
\label{fig:PM_quench_0.1} 
\end{figure*}

\subsection{Quenches from the paramagnetic phase}
As we shall see in the following, the DPT-I and DPT-II dynamical critical lines also coincide when starting with paramagnetic thermal initial states, albeit their shape will be qualitatively different from the case of ferromagnetic initial conditions.

For the system to be able to detect the possible preference for ferromagnetic order in the final state following a purely unitary time evolution, we have to introduce a finite seed in the form of a magnetic field along the $z$-direction with strength $\Lambda >0$. This is not a finite-size effect and even the thermodynamic system will not exhibit spontaneous symmetry breaking. Instead the inability of the system to dissipate energy forces the final magnetization in the ferromagnetic phase to depend on the initial magnetization. Given an initial state with $\langle S_z\rangle = \langle S_y\rangle = 0$, the thermodynamic system will not show any dynamics at all, independent of the final value $\Gamma_\text{f}$. Therefore, for all the plots presented in Figs.~\ref{fig:PM_quench_0.6},~\ref{fig:PM_quench_0.2}, and~\ref{fig:PM_quench_0.1}, we set $\Lambda_{\text{i}}=\Gamma_{\text{i}}/20$. The motion of the magnetization will be determined by the angles $\bar{\theta}_\text{i}$ and $\bar{\theta}_\text{f}$ that minimize the pre- and post-quench classical Hamilton function. This angular dependence leads to the following consequence: Every finite difference $\bar{\theta}_\text{i}-\bar{\theta}_\text{f}$ gives rise to a non-stationary magnetization, \textit{e.g.}~we can set $\Lambda_{\text{f}}=0$, which for quenches where the final state is still paramagnetic results in the long-time average $\langle S_z\rangle =0$ in contrast to $\langle S_z\rangle \neq 0$ for quenches to a ferromagnetic state. On the other hand, whenever the classical spin expectation value moves, we find a return function that does not scale to zero in the thermodynamic limit. Within the paramagnetic phase such a classical motion of the total magnetization vector, which is purely caused by the need of a small explicit symmetry breaking in the initial state, can be avoided by choosing the final external field $\Lambda_\text{f}$ such that the angles $\bar{\theta}_{{\text{i}},{\text{f}}}$, coincide. The absence of classical motion yields an entirely smooth return rate that scales to zero in the thermodynamic limit. Quenches that remain in the paramagnetic phase in the DPT-I sense can therefore be classified as \textit{trivial} in the DPT-II sense. In general, the resulting trigonometric equation for $\Lambda_\text{f}$ has to be solved numerically, as is done for Figs.~\ref{fig:PM_quench_0.6},~\ref{fig:PM_quench_0.2}, and~\ref{fig:PM_quench_0.1}. For small values of $\Lambda_{{\text{i}},{\text{f}}}/\Gamma_{{\text{i}},{\text{f}}}$, however, this reduces to the simple expression
\begin{align}\label{eq:hf}
\Lambda_{\text{f}}=\Lambda_{\text{i}}\frac{\bar{s}-2\Gamma_{\text{f}}}{\bar{s}-2\Gamma_{\text{i}}}\;.
\end{align}
The important difference for a quench to the ferromagnetic phase lies in that \eqref{eq:hf} has no solution and even setting $\Lambda_{\text{f}}=0$ results in a final magnetization along the $z$-direction that is larger than that of the initial state. The unavoidable classical motion of the magnetization vector results in a finite return rate with regular cusps as $N\to\infty$, which indicates the same dynamical phase transition as characterized by the DPT-I. Apart from the relaxation due to dephasing, the time evolution and mean value of the magnetization are again well-described by the classical equations of motion.

In Fig.~\ref{fig:PM_quench_0.6}, we show ED results for a paramagnetic thermal initial state at $\Gamma_\text{i}=1$ that is subsequently subjected to a quench in the transverse-field strength to the value $\Gamma_\text{f}=0.6$. Fig.~\ref{fig:PM_quench_0.6} shows this quench where the initial states are prepared at $T=0.2$, $T=4/3$ and $T=10$. Each of the return rates shows a trivial phase\cite{Heyl2013,Halimeh2016b} and scales to zero in the thermodynamic limit. Since we break the $\mathbb{Z}_2$ symmetry explicitly by a small external magnetic field along the $z$ direction in both the initial and final Hamiltonian, with the value of $\Lambda_\text{f}$ chosen appropriately, the initial and final magnetization after the decay of the induced oscillations are the same. This value of the magnetization in the thermodynamic limit can be easily found from the classical model introduced in Sec.~\ref{sec:classical}.

Fig.~\ref{fig:PM_quench_0.2} shows the same analysis for the quench from $\Gamma_\text{i}=1\to\Gamma_\text{f}=0.2$. At a sufficiently low temperature $T=0.2$, we see in Fig.~\ref{fig:PM_quench_0.2}(a) that the dynamics gives rise to a ferromagnetic steady state with infinite-time average of the magnetization greater than the seeding value. This ordered infinite-time steady state coincides with a regular phase in the return rate characterized by a cusp in each period of $r_\text{q}(t)$. Corresponding insets show how the curvature of $r_\text{q}(t)$ at the first cusp diverges algebraically with system size, indicating clear non-analytic behavior in the thermodynamic limit. Upon further increasing the temperature to $T=4/3$ or even $T=10$, the dynamics no longer leads to an ordered steady state and the regular phase is replaced by the trivial phase that goes to zero in the thermodynamic limit, as shown in Fig.~\ref{fig:PM_quench_0.2}(b,c).

Fig.~\ref{fig:PM_quench_0.1} repeats this analysis but at an even larger quench from $\Gamma_\text{i}=1\to\Gamma_\text{f}=0.1$. While the magnetization and return rate for panels (a) and (c) are qualitatively similar to the corresponding temperatures in Fig.~\ref{fig:PM_quench_0.2}, the regular phase now also replaces the trivial quench at $T=4/3$. The finite-size scaling in Fig.~\ref{fig:PM_quench_0.1}(b), on which we base this claim, requires much larger systems than the other quenches. This is because at high temperatures and close to the critical field strength the dynamics governing the system is slow and fluctuations that introduce dephasing are enhanced. The combination of both lead to unusually strong finite-size effects. Consequently, both the magnetization and return rate converge much more slowly towards the thermodynamic limit. At first sight, a regular phase for this quench is surprising since in equilibrium there is no ferromagnetic phase at these high temperatures. However, the conserved spin length $S$ for these quenches starting from deep within the paramagnetic phase is longer than the equilibrium value at $\Gamma_\text{f}$, which in turn increases the system's susceptibility to ferromagnetic order. Indeed, we find, that no matter how high the temperature of the thermal initial state is, there is always a small enough $\Gamma_\text{f}$ to which a quench would give rise to a ferromagnetic infinite-time steady state that coincides with a regular phase. For a system that is weakly connected by local couplings to its environment this has the interesting consequence that there exists a timescale in the relaxation from a paramagnetic thermal initial state to a paramagnetic long-time steady state during which the system can spontaneously break $\mathbb{Z}_2$ symmetry and thus evolve through a ferromagnetic quasi-stationary state. Here, energy dissipation due to the local contact to the environment allows for relaxation of the length of the magnetization vector. Consequently, this enables the system to evolve from a ferromagnetic state, which would be the infinite-time steady state in the case of a closed system, to a paramagnetic equilibrium final state.

Also in the case of quenches from a paramagnetic thermal state, the return rate and the magnetization profile exhibit the same periodic relation as has been shown in the literature. For quenches that end up in the paramagnetic phase, the largest deviation between $\langle m(t) \rangle$ and $\langle m(0) \rangle$ coincides with times when $\langle S_z(t) \rangle$ takes its initial value; however, the deviation in the azimuthal angle becomes maximal.

\begin{figure}\centering
\includegraphics[width=0.95 \columnwidth]{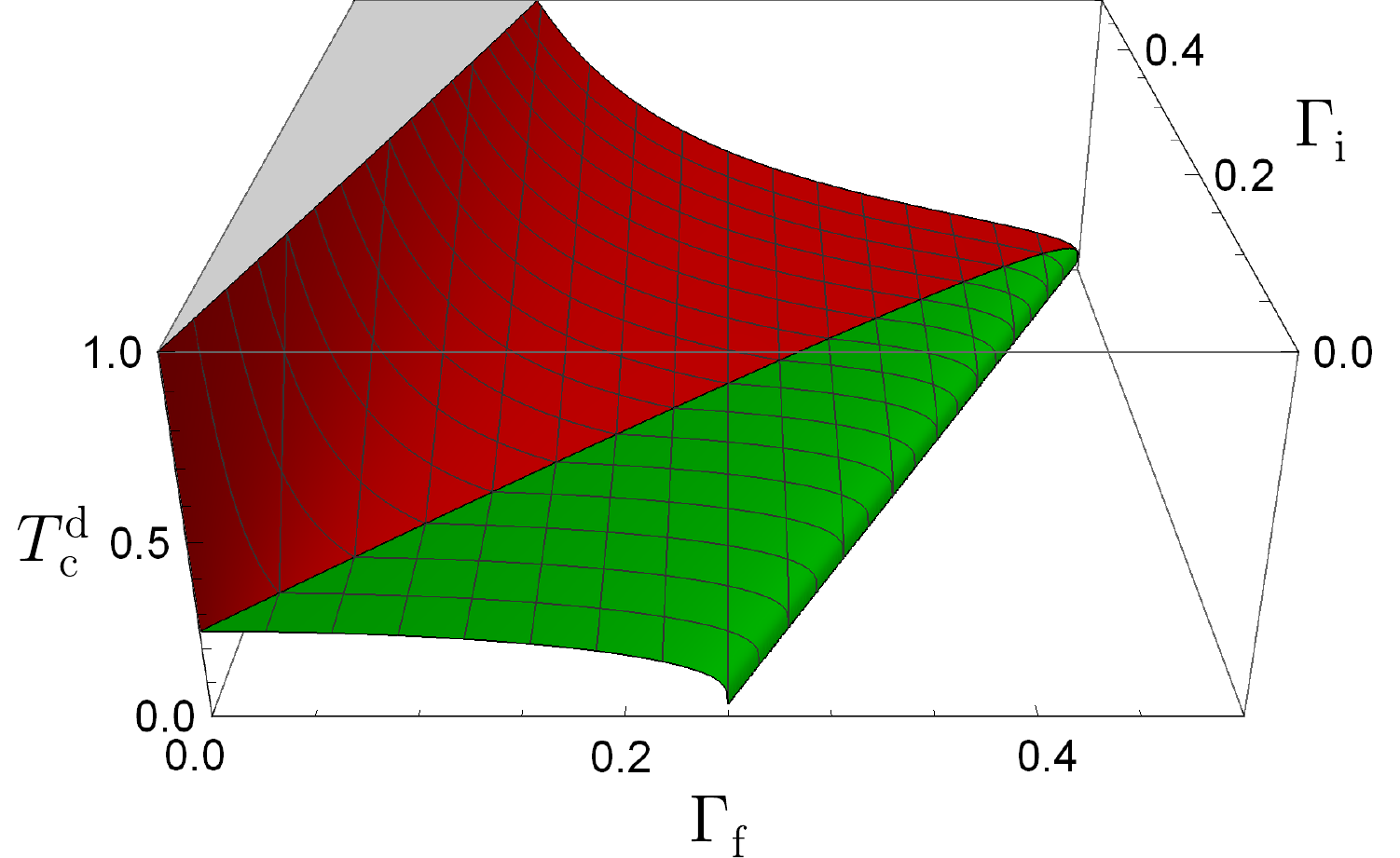}
\caption{(Color online) Finite-temperature dynamical phase diagram of the fully connected transverse-field Ising model for ferromagnetic (green) and paramagnetic (red) initial states. Analytical results from~\eqref{eq:T_c_dyn} and~\eqref{eq:G_c_dyn} coincide with numerical results for the return rate, for which no error bars are shown, since they are in most cases too small to be resolved in the plot.}
\label{fig:ana_phase}
\end{figure}

Within our numerical precision, we find from our ED simulations that for quenches beginning from a paramagnetic thermal state an ordered (a disordered) infinite-time steady state always coincides with a regular (trivial) phase in the return rate~\eqref{eq:rtilde}, and thus as for quenches from the ordered phase, the DPT-I and DPT-II share the same critical line. Unlike for quenches from a ferromagnetic thermal initial state, the dynamical critical line here cannot be directly connected to its equilibrium counterpart. Finally, we summarize our findings with regards to the dynamical critical line for all initial conditions in Fig.~\ref{fig:ana_phase}.

\section{Conclusion}\label{sec:conclusion}
We have studied two types of dynamical phase transitions in the fully connected transverse-field Ising model that arise upon quenching a ferromagnetic or paramagnetic thermal initial state. Whereas one type of dynamical phase transition is characterized by the long-time average of the local $\mathbb{Z}_2$ order parameter, the second is characterized by the existence and type of cusps in a properly defined Loschmidt-echo return rate that captures non-analyticities due to only the quantum quench dynamics and suppresses interferences between different thermally occupied spin subspaces. In agreement with what is known at zero temperature,\cite{Homrighausen2017} starting in an ordered thermal initial state and quenching below a dynamical critical point leads to a phase that is also ordered in the long-time limit in the Landau sense, and where the corresponding return rate exhibits anomalous cusps, which appear only after the first minimum of the return rate. The presence of an anomalous phase is intimately connected with the long-ranged character of the interactions. These suppress fluctuations, and in an effective field theory picture allow the order-parameter field to maintain its cohesion through at least one period. If the interaction exponent is increased above a quench protocol-dependent critical value $\alpha_c$ of roughly $2.3-2.4$, the dominant type of excitation changes in character from single-spin flips to domain walls.\cite{Vanderstraeten2018} These make the order-parameter field disperse faster and thereby destroy the anomalous phase.

In the case of the ordered thermal state undergoing a quench to above the dynamical critical point, where the order-parameter evolution explores both hemispheres of the Bloch sphere, the long-time steady state will be disordered and the return rate will show regular cusps. The critical line shared by these two types of dynamical phase transition is dependent on the initial conditions (temperature and transverse-field strength) within the ordered phase, and can be directly connected to the equilibrium critical line.

As for quenches starting from a disordered thermal initial state, a small quench distance leads to a trivial phase, in which the return rate goes to zero in the thermodynamic limit, and which concurs with a disordered infinite-time steady state. However, for a large enough quench distance towards smaller transverse fields the return rate will exhibit a regular phase and the infinite-time steady state will carry ferromagnetic order, regardless at what temperature the paramagnetic thermal initial state is prepared. Thus, even though both types of dynamical criticality also share a common critical line for quenches from paramagnetic thermal initial states, this dynamical critical line cannot be directly connected to the equilibrium critical line.

\section*{Acknowledgments}
We are grateful to Amit Dutta, Markus Heyl, Ingo Homrighausen, and Francesco Piazza for stimulating discussions. This project has been supported by NIM (Nano Initiative Munich).

\appendix

\begin{figure*}[htp]
\centering
\hspace{-.25 cm}
\includegraphics[width=.335\textwidth]{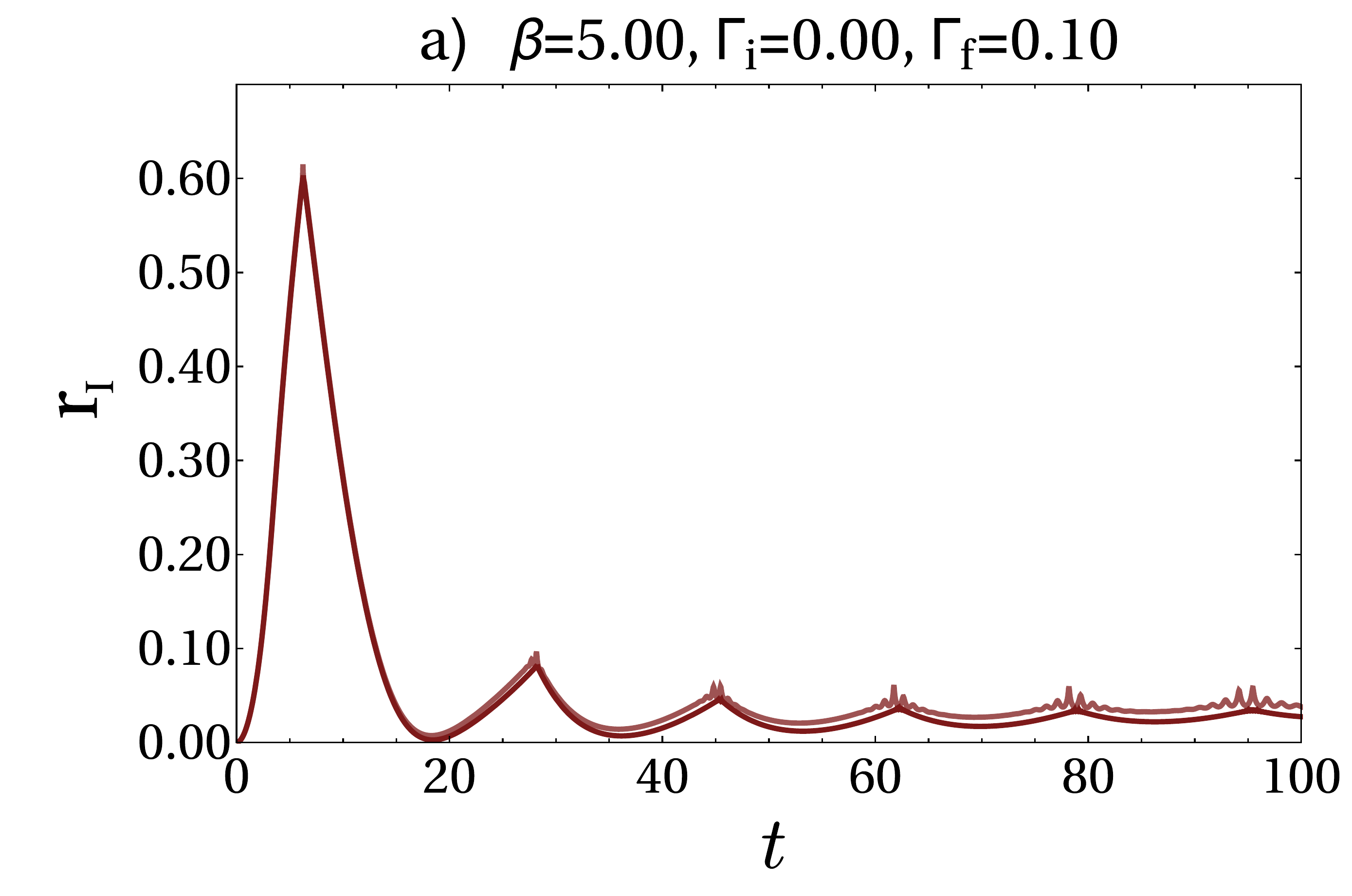}\quad
\hspace{-.4 cm}
\includegraphics[width=.335\textwidth]{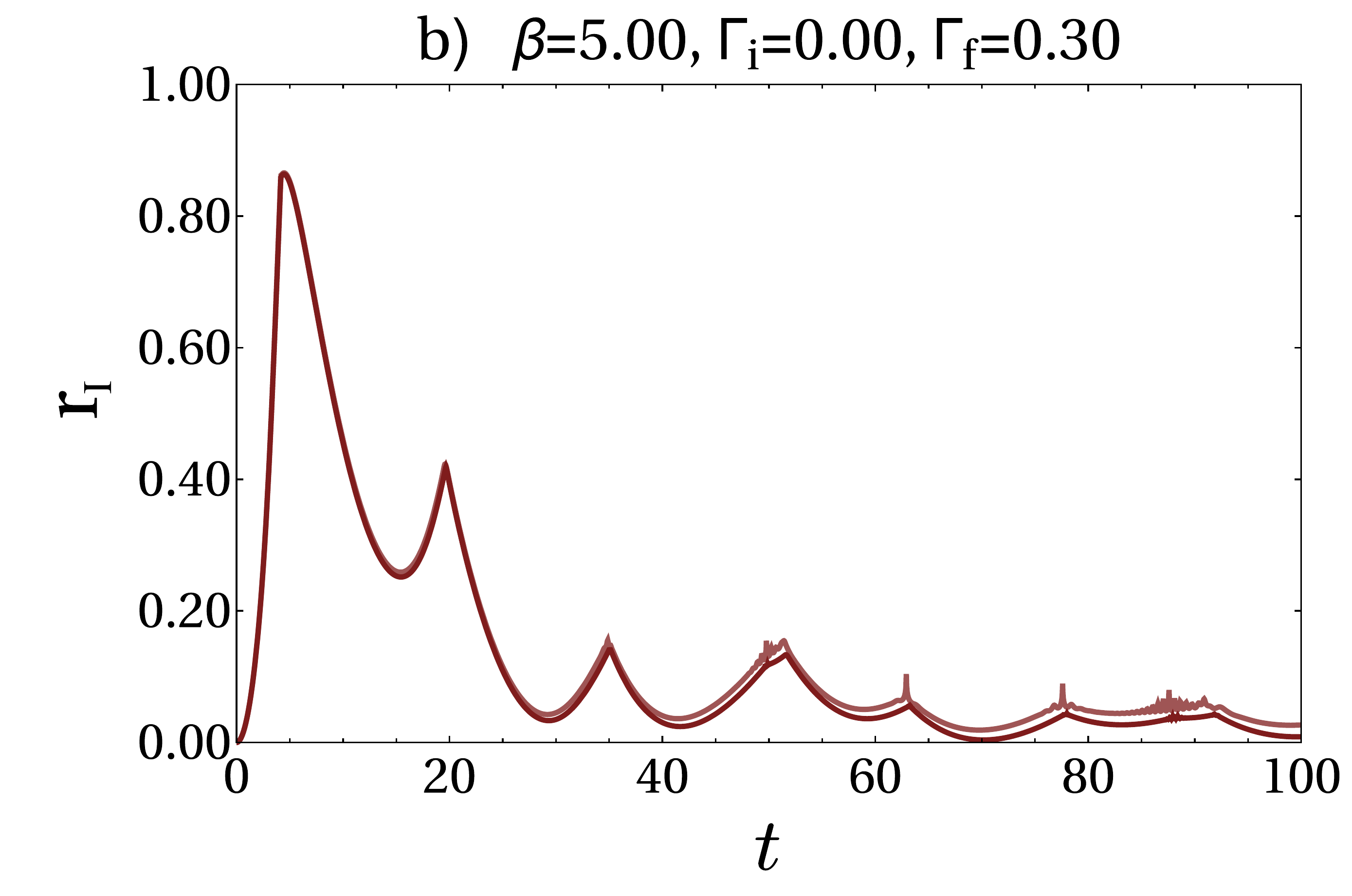}\quad
\hspace{-.4 cm}
\includegraphics[width=.335\textwidth]{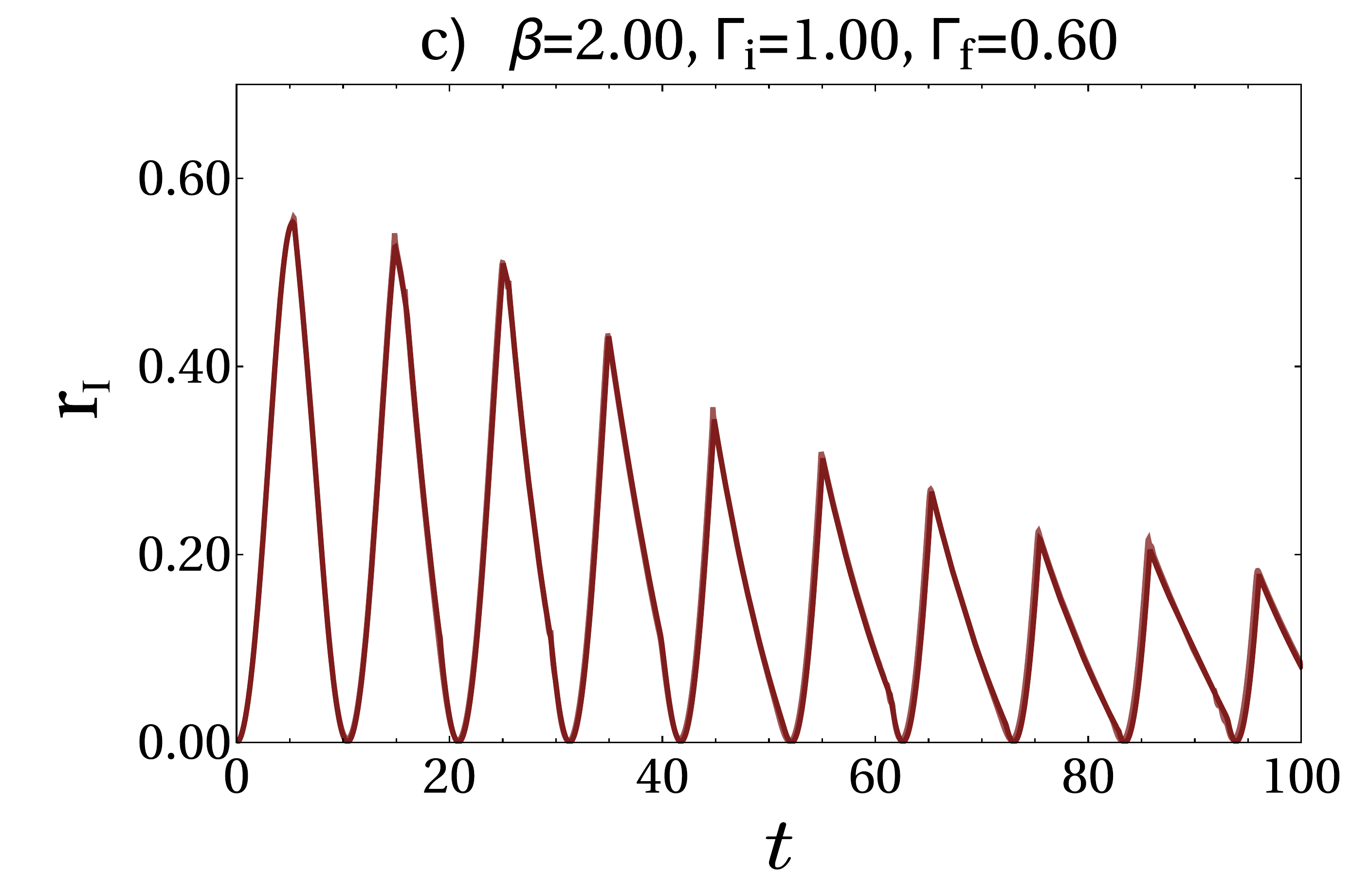}
\hspace{-.15 cm}
\caption{(Color online) Behavior of the interferometric return rate $r_\text{I}(t)$ as defined in \eqref{eq:LERR_thermal} for different quenches, which also shown in the main text for $r_\text{q}(t)$, with system sizes $N=201$ and $N=1001$ for the light and dark line respectively. The quench in panel (a) is identical to the one depicted in Fig.~\ref{fig:FM_quench_0.1}(b). Here $r_\text{I}(t)$ exhibits pronounced non-analyticities in all peaks. The same is true for the quench in (b), which is the same as in Fig.~\ref{fig:FM_quench_0.3}(b), and for the quench in panel (c) that is deep within the paramagnetic phase and also shown in Fig.~\ref{fig:PM_quench_0.6}(b) for $r_\text{q}(t)$.}
\label{fig:R} 
\end{figure*}
\begin{figure*}[htp]
\centering
\hspace{-.25 cm}
\includegraphics[width=.335\textwidth]{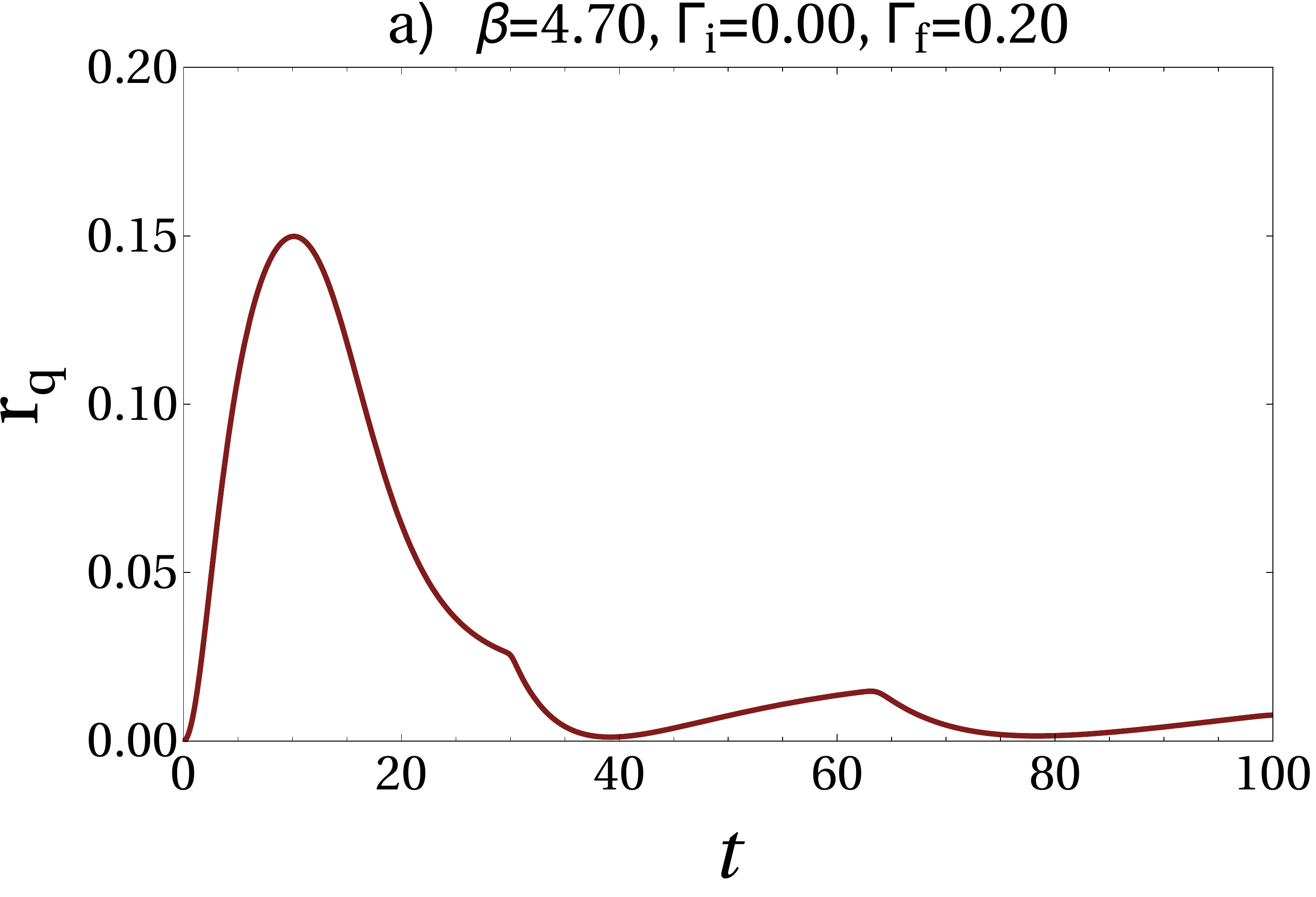}\quad
\hspace{-.4 cm}
\includegraphics[width=.335\textwidth]{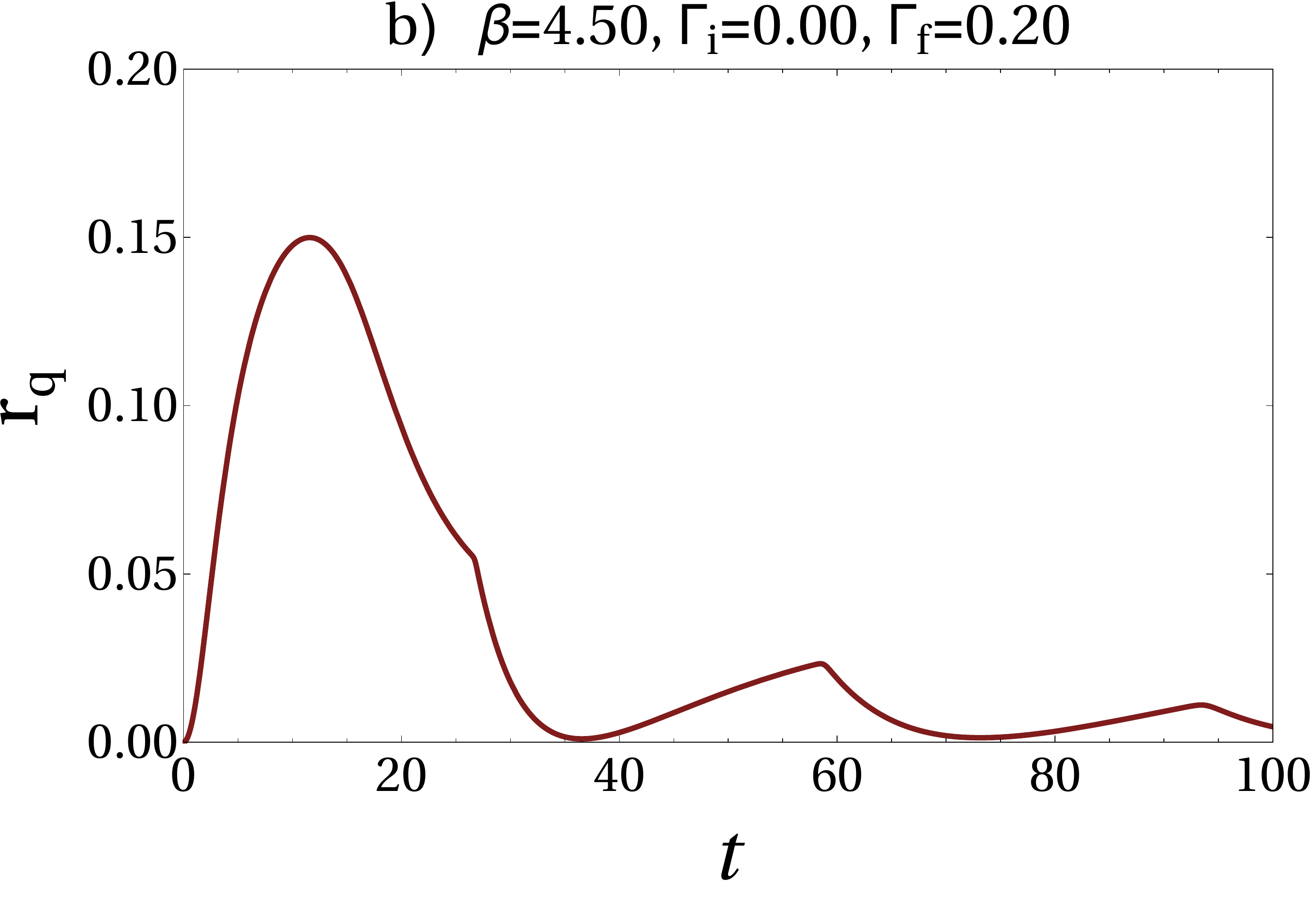}\quad
\hspace{-.4 cm}
\includegraphics[width=.335\textwidth]{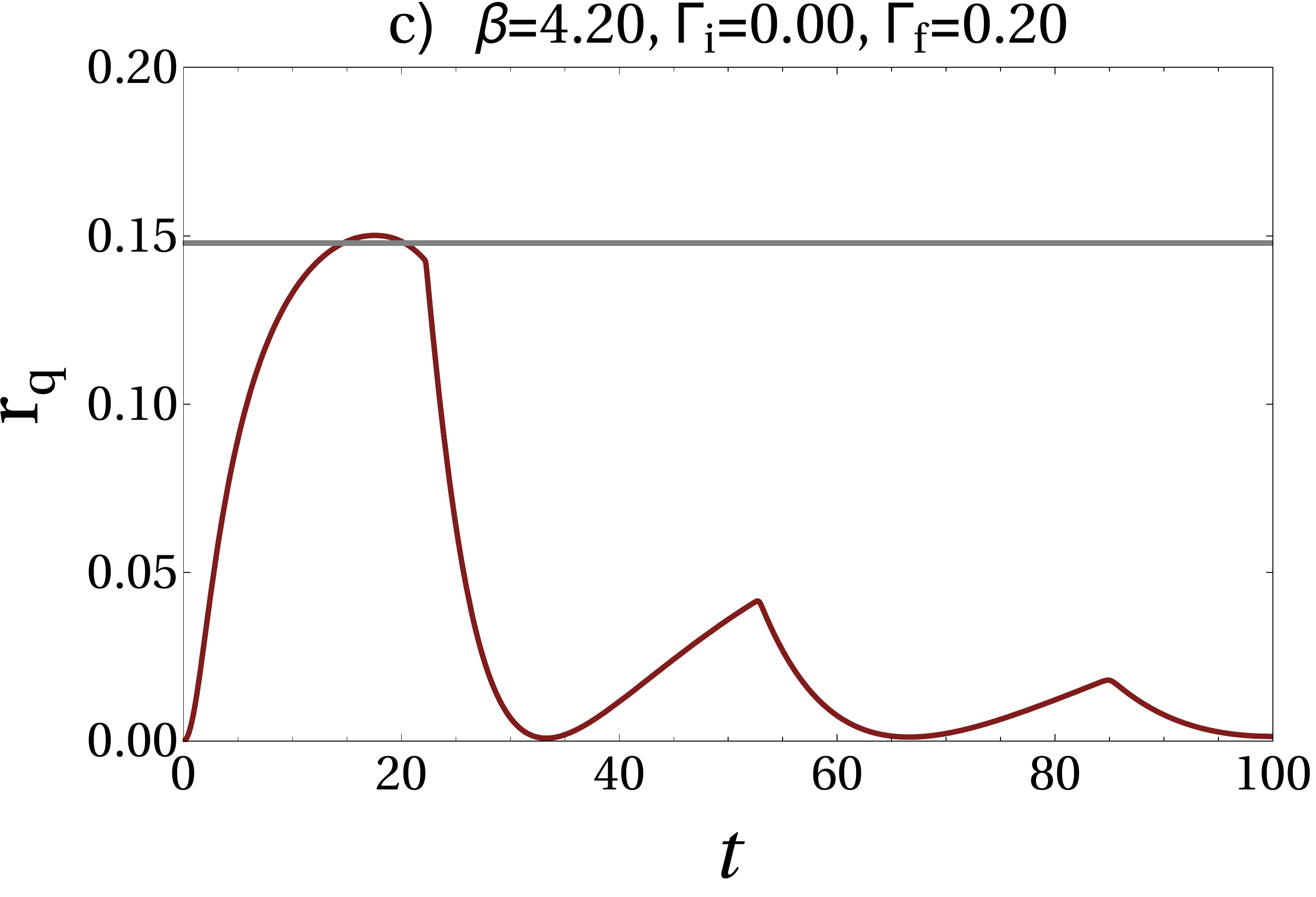}
\hspace{-.15 cm}
\caption{(Color online) Disappearance of the first cusp underneath the thermal cutoff with increasing temperature. From left to right the temperature grows form $\beta=4.7$ through $\beta=4.5$ in panel (b) to $\beta=4.2$. While these quenches reach deeper and deeper into the regular phase the simultaneously decreasing value of the thermal cutoff will eventually crop the first cusp. The constant indicated by the gray line in c) represents $r_\text{q}^{\text{max}}$ as given by \eqref{eq:rmax}, thus the signal will be cut off in the thermodynamic limit. System size is $N=2001$.} 
\label{fig:Hide} 
\end{figure*}

\section{Analytics of the interferometric return rate}
\label{app:thermal_ana}
In the absence of a transverse field, the Hamiltonian of the fully connected Ising model reduces to~\eqref{eq:NoTF} with degeneracies 

\begingroup
\renewcommand{\arraystretch}{1.5}
\begin{align}
D_z(S_z)=
\begin{pmatrix}
	N \\
	S_z+\frac{N}{2}
\end{pmatrix}.
\end{align}
\endgroup
While the return rate, even for this simple system, cannot be calculated exactly for arbitrary system sizes $N$, we can obtain the thermodynamic limit for short times analytically. For large systems and short times, the sum in the return rate for $\Gamma_\text{i}=\Gamma_\text{f}$, which is given by

\begin{align}\label{eq:trivialQuenchExact}
r_\text{I}(t)=-\frac{2}{N}\ln\left|\sum_{S_z} D_z(S_z)\exp\left[\left(\beta+\text{i}t\right)\frac{S_z^2}{2N}\right]\right|+\frac{2}{N} \ln Z,
\end{align}
can be replaced by an integral. With $s_z=\frac{S_z}{N}+\frac{1}{2}$ we obtain
\begin{align}
r_\text{I}(t)=&-\frac{2}{N}\ln\bigg|\int_0^{1/2}\d s_z\;\exp \bigg\{N\bigg[\left(\beta+\text{i}t\right)\frac{1}{8}(1-2s_z)^2\notag \\\label{eq:r_int}
&-s_z\ln{s_z}-(1-s_z)\ln(1-s_z)-\ln{2}\bigg]\bigg\}\bigg|+A\;,
\end{align}
 to leading order in $N$, where the constant $A$ ensures the normalization $r(t=0)=0$. Its value $A=\frac{1}{4}\left[\beta-4s_0^2\beta-8\ln{\left(2-2 s_0\right)}\right]$ is determined by the evaluation of the integrand in \eqref{eq:r_int} at the non-trivial saddle point $s_0 \in \left(0, 1/2 \right)$, which solves 
\begin{align}
\beta\left(\frac{1}{2}-s_0\right)=2\arctanh{\left(1-2s_0\right)}\;.
\end{align} 
For finite $t>0$, however, the first term of the return rate $r_\text{I}(t)$ will be dominated by the values of $s_z$ near $1/2$, where the exponent converges quadratically to zero. Yet, in the thermodynamic limit the integral over this region yields only a vanishing contribution to $r_\text{I}(t)$ such that the return rate is bounded by $A$ from above. 
For short times in the sense of $t \cdot \Delta \epsilon \ll 1$, where $\Delta \epsilon$ is the typical energy difference between the discrete levels around the saddle point $s_0$, one can still use~\eqref{eq:r_int} as an approximation to $r_\text{I}(t)$. Performing again a saddle point expansion around $s_0$, the ensuing Gaussian integral and the limit $N \to \infty $ yields
\begin{align}\label{eq:r_gauss}
r_\text{I}(t)=\frac{s_0\left(1-s_0\right)\left(1-2 s_0\right)^2\left[1+s_0\left(s_0-1\right)\beta\right]t^2}{4\left\{1+s_0\left(s_0-1\right)\left[2\beta+s_0\left(s_0-1\right)\left(t^2+\beta^2\right)\right]\right\}}\;.
\end{align}
By comparing the result for $A$ and \eqref{eq:r_gauss} we realize that for sufficiently small temperatures the unrestricted Gaussian integral used in \eqref{eq:r_gauss} allows $r_\text{I}(t)$ to quickly grow beyond its bound set by $A$. This is not possible for the original expression \eqref{eq:trivialQuenchExact} or \eqref{eq:r_int}, involving a \emph{restricted} sum or integral instead. Consequently, one has a sharp transition from the short time behavior to the limiting value.

\section{Numerical results for the interferometric return rate}
\label{app:thermal_num}
We have already shown in the main text that, even for trivial quenches, $r_\text{I}(t)$ can still exhibit anomalous cusps at finite temperatures that in the usual case at $T=0$ would be interpreted as an indication of a dynamical phase transition. Here we will give a few examples of how the behavior of this return rate gets even more convoluted for finite quench distances.

The anomalous behavior of $r_\text{I}(t)$ for trivial quenches quickly turns into a regular behavior with very pronounced cusps in every peak for short quench distances beginning and ending in an ordered state. An example for the same parameters as in Fig.~\ref{fig:FM_quench_0.1}(b) is illustrated in Fig.~\ref{fig:R}(a). Interestingly, while $r_\text{q}(t)$ is a lower bound on $r_\text{I}(t)$ it is only tight at extremely low temperatures ($T\ll0.1$) and thus no conclusions based on the behavior of $r_\text{q}(t)$ can be drawn for the interferometric return rate.

This complete independence between $r_\text{I}(t)$ on the one hand and $r_\text{q}(t)$ and the DPT-I on the other hand is revealed again for quenches through the dynamical phase transition. As a representative example we present the same quench as in Fig.~\ref{fig:FM_quench_0.3}(b) in Fig.~\ref{fig:R}(b). While there are clearly more cusps visible in Fig.~\ref{fig:R}(b) than in Fig.~\ref{fig:R}(a) these additional cusps do not appear all at the same time and cannot be linked to any particular behavior of the magnetization vector.

Finally, we note that even a quench at very high temperature and very deep within the disordered phase, as shown in Fig.~\ref{fig:R}(c), shows cusps in every peak of $r_\text{I}(t)$. This happens despite the absence of any motion in the magnetization vector and in a regime where not even $T=0$ would support ferromagnetic order.\\

\section{Thermal cutoff and the regular phase}
\label{app:illustration}
In Fig.~\ref{fig:Hide} we present an example of the emergence of a thermal cutoff in the regular phase. In particular, this illustrates why we do not attribute this behavior to a dynamical phase transition. Beginning at a temperature $\beta=4.7$ we quench from $\Gamma_\text{i}=0$ to $\Gamma_\text{f}=0.2$, obtaining regular behavior where the first cusp appears shortly before the first minimum at finite time. Increasing the temperature to $\beta=4.5$, the cusp appears earlier in time at a larger value of the return rate. Finally, at $\beta=4.2$ the thermal cutoff is almost small enough to affect the first peak and the cusp that is now located almost at the top of the peak. Actually, in the thermodynamic limit the cutoff will be small enough to affect the very top of the first maximum of the return rate. However for the system size shown in Fig.~\ref{fig:Hide}(c) this is not yet the case.


\bigskip

\begin{thebibliography}{9}
\bibitem{Ma}
S.-K. Ma, \textit{Statistical Mechanics} (World Scientific Publishing, 1985).
\bibitem{Cardy}
J. Cardy, \textit{Scaling and Renormalization in Statistical Physics} (Cambridge University Press, 1996).

\bibitem{Sachdev}
S. Sachdev, \textit{Quantum Phase Transitions} (Cambridge University Press, 1999).

\bibitem{Fisher1974}
M. Fisher, Rev. Mod. Phys. \textbf{46}, 597 (1974).

\bibitem{Greiner2002}
M. Greiner, O. Mandel, T. Esslinger, T. W. H\"ansch, and I. Bloch, Nature \textbf{415}, 39 (2002).

\bibitem{Bloch2008}
I. Bloch, J. Dalibard, and W. Zwerger, Rev. Mod. Phys. \textbf{80}, 885 (2008).

\bibitem{Hart2014}
S. Hart \textit{et al.}, Nat. Phys. \textbf{10}, 638 (2014). 

\bibitem{Suda2015}
M. Suda \textit{et al.}, Science \textbf{347}, 743 (2015).

\bibitem{Mitrano2016}
M. Mitrano \textit{et al.}, Nature \textbf{530}, 461 (2016).

\bibitem{Zvyagin2017}
A. A. Zvyagin, Low Temperature Physics \textbf{42}, 971 (2016).

\bibitem{Moeckel2008}
M. Moeckel and S. Kehrein, Phys. Rev. Lett. \textbf{100}, 175702 (2008).

\bibitem{Eckstein2008}
M. Eckstein and M. Kollar, Phys. Rev. Lett. \textbf{100}, 120404 (2008).

\bibitem{Eckstein2009}
M. Eckstein, M. Kollar, and P. Werner, Phys. Rev. Lett. \textbf{103}, 056403 (2009).

\bibitem{Moeckel2010}
M. Moeckel and S. Kehrein, New J. Phys. \textbf{12}, 055016 (2010).

\bibitem{Sciolla2010}
B. Sciolla and G. Biroli, Phys. Rev. Lett. \textbf{105}, 220401 (2010).

\bibitem{Sciolla2011}
B. Sciolla and G. Biroli, J. Stat. Mech. (2011) P11003.

\bibitem{Gambassi2011}
A. Gambassi and P. Calabrese, Europhys. Lett. \textbf{95}, 66007 (2011).

\bibitem{Tsuji2013}
N. Tsuji, M. Eckstein, and P. Werner, Phys. Rev. Lett. \textbf{110}, 136404 (2013).
 
\bibitem{Sciolla2013}
B. Sciolla and G. Biroli, Phys. Rev. B \textbf{88}, 201110 (2013).

\bibitem{Chandran2013}
A. Chandran, A. Nanduri, S. S. Gubser, and S. L. Sondhi, Phys. Rev. B \textbf{88}, 024306 (2013).

\bibitem{Maraga2015}
A. Maraga, A. Chiocchetta, A. Mitra, and A. Gambassi, Phys. Rev. E \textbf{92}, 042151 (2015).

\bibitem{Smacchia2015}
P. Smacchia, M. Knap, E. Demler, and A. Silva, 
Phys. Rev. B \textbf{91}, 205136 (2015).

\bibitem{Langen2016}
T. Langen, T. Gasenzer, and J. Schmiedmayer, J. Stat. Mech. (2016) 064009.

\bibitem{Marcuzzi2016}
M. Marcuzzi, J. Marino, A. Gambassi, and A. Silva, Phys. Rev. B \textbf{94}, 214304 (2016).

\bibitem{Zunkovic2016a}
B. Zunkovic, A. Silva, and M. Fabrizio, 
Phil. Trans. R. Soc. A \textbf{374}, 20150160 (2016).

\bibitem{Halimeh2016a}
J. C. Halimeh, V. Zauner-Stauber, I. P. McCulloch, I. de Vega, U. Schollw\"ock, and M. Kastner, Phys. Rev. B \textbf{95}, 024302 (2017).

\bibitem{Halimeh2016b}
J. C. Halimeh and V. Zauner-Stauber, Phys. Rev. B \textbf{96}, 134427 (2017).

\bibitem{Homrighausen2017}
I. Homrighausen, N. O. Abeling, V. Zauner-Stauber, and J. C. Halimeh, Phys. Rev. B \textbf{96}, 104436 (2017).

\bibitem{Halimeh2017a}
V. Zauner-Stauber and J. C. Halimeh, Phys. Rev. E \textbf{96}, 062118 (2017).

\bibitem{Zunkovic2016b}
B. Zunkovic, M. Heyl, M. Knap, and A. Silva, Phys. Rev. Lett. \textbf{120}, 130601 (2018).

\bibitem{Zhang2017}
J. Zhang \textit{et al.}, Nature \textbf{551}, 601 (2017).

\bibitem{Mori2017}
T. Mori, T. N. Ikeda, E. Kaminishi, and M. Ueda, arXiv:1712.08790.

\bibitem{Heyl2013}
M. Heyl, A. Polkovnikov, and S. Kehrein, Phys. Rev. Lett. \textbf{110}, 135704 (2013).

\bibitem{Karrasch2013}
C. Karrasch and D. Schuricht, Phys. Rev. B \textbf{87}, 195104 (2013).

\bibitem{Andraschko2014}
F. Andraschko and J. Sirker, 
Phys. Rev. B \textbf{89}, 125120 (2014).

\bibitem{Vajna2014}
S. Vajna and B. D\'ora,
Phys. Rev. B \textbf{89}, 161105(R) (2014).

\bibitem{Heyl2014}
M. Heyl, Phys. Rev. Lett. \textbf{113}, 205701 (2014).

\bibitem{Heyl2015}
M. Heyl, 
Phys. Rev. Lett. \textbf{115}, 140602 (2015).

\bibitem{Budich2016}
J. C. Budich and M. Heyl, Phys. Rev. B \textbf{93}, 085416 (2016).

\bibitem{Divakaran2016}
U. Divakaran, S. Sharma, and A. Dutta, Phys. Rev. E \textbf{93}, 052133 (2016).

\bibitem{Abeling2016}
N. O. Abeling and S. Kehrein, 
Phys. Rev. B \textbf{93}, 104302 (2016).

\bibitem{Karrasch2017}
C. Karrasch and D. Schuricht, Phys. Rev. B \textbf{95}, 075143 (2017).

\bibitem{Weidinger2017}
S. A. Weidinger, M. Heyl, A. Silva, M. Knap, Phys. Rev. B \textbf{96}, 134313 (2017).

\bibitem{Dutta2017}
U. Bhattacharya, S. Bandopadhyay, and A. Dutta, 	Phys. Rev. B \textbf{96}, 180303(R) (2017).

\bibitem{Heyl2017}
M. Heyl and J. C. Budich, Phys. Rev. B \textbf{96}, 180304(R) (2017).

\bibitem{Sirker2017}
N. Sedlmayr, M. Fleischhauer, and J. Sirker, Phys. Rev. B \textbf{97}, 045147 (2018).

\bibitem{Zhou2017}
L. Zhou, Q.-h. Wang, H. Wang, and J. Gong, arXiv:1711.10741.

\bibitem{Mera2017}
B. Mera, C. Vlachou, N. Paunkovi\'c, V. R. Vieira, and O. Viyuela, Phys. Rev. B \textbf{97}, 094110 (2018).

\bibitem{Flaeschner2016}
N. Fl\"aschner \textit{et al.}, Nat. Phys. (2017).

\bibitem{Jurcevic2017}
P. Jurcevic \textit{et al.}, Phys. Rev. Lett. \textbf{119}, 080501 (2017).

\bibitem{HeylReview}
M. Heyl, arXiv:1709.07461.

\bibitem{Wang2017}
Q. Wang and H. T. Quan, Phys. Rev. E \textbf{96}, 032142 (2017).
 
\bibitem{Kac1963}
M. Kac, G. E. Uhlenbeck, and P. C. Hemmer, J. Math. Physics \textbf{4}, 216 (1963).

\bibitem{Das2006}
A. Das, K. Sengupta, D. Sen, and B. Chakrabarti, Phys. Rev. B \textbf{74}, 144423 (2006).

\bibitem{Wilms2011}
J. Wilms, J. Vidal, F. Verstraete, and S. Dusuel, J. Stat. Mech. (2012) P01023.

\bibitem{Lipkin1965}
H. J. Lipkin, N. Meshkov, and A. J. Glick, Nucl. Phys. \textbf{62}, 188 (1965).

\bibitem{Meshkov1965}
N. Meshkov, A. J. Glick, and H. J. Lipkin, Nucl. Phys. \textbf{62}, 199 (1965).

\bibitem{Glick1965}
A. J. Glick, H. J. Lipkin, and N. Meshkov, Nucl. Phys. \textbf{62}, 211 (1965).

\bibitem{Botet1982}
R. Botet, R. Jullien, and P. Pfeuty, Phys. Rev. Lett. \textbf{49}, 478 (1982).

\bibitem{Botet1983}
R. Botet and R. Jullien, Phys. Rev. B \textbf{28}, 3955 (1983).

\bibitem{Morrison2008}
S. Morrison and A. S. Parkins, Phys. Rev. Lett. \textbf{100}, 040403 (2008).

\bibitem{Hess2017}
P. W. Hess \textit{et al.}, Phil. Trans. R. Soc. A \textbf{375}, 0107 (2017).

\bibitem{Bernien2017}
H. Bernien \textit{et al}, Nature \textbf{551}, 579 (2017).

\bibitem{Neyenhuis2017}
B. Neyenhuis \textit{et al.}, Science Advances \textbf{3}, e1700672 (2017).

\bibitem{Ribeiro2008}
P. Ribeiro, J. Vidal, and R. Mosseri, Phys. Rev. E \textbf{78}, 021106 (2008).

\bibitem{Lee1952}
T. D. Lee and C. N. Yang Phys. Rev. \textbf{87}, 410 (1952).

\bibitem{Zanardi2007}
P. Zanardi, H. T. Quan, X. Wang, and C. P. Sun, Phys. Rev. A \textbf{75}, 032109 (2007).

\bibitem{Venuti2011}
L. C. Venuti, N. T. Jacobson, S. Santra, and P. Zanardi, Phys. Rev. Lett. \textbf{107}, 010403 (2011).

\bibitem{Vanderstraeten2018}
L. Vanderstraeten, M. Van Damme, H. P. B\"uchler, and F. Verstraete, arXiv:1801.00769.

\end{thebibliography}
\end{document}